\documentclass[lettersize,journal]{IEEEtran}
\usepackage{amsmath,amsfonts}
\usepackage{algorithmic}
\usepackage{algorithm}
\usepackage{array}
\usepackage[caption=false,font=normalsize,labelfont=sf,textfont=sf]{subfig}
\usepackage{textcomp}
\usepackage{stfloats}
\usepackage{url}
\usepackage{verbatim}
\usepackage{graphicx}
\usepackage[percent]{overpic}
\usepackage{xcolor}
\usepackage{cite}
\usepackage{hyperref}
\newcommand{\vect}[1]{\boldsymbol{#1}}

\begin{document}

\title{DynamicSound simulator for simulating moving sources and microphone arrays}

\author{Luca Barbisan, Marco Levorato, Fabrizio Riente
\thanks{Luca Barbisan and Fabrizio Riente are with the Department of Electronics and Telecommunications, Politecnico di Torino, Torino, Italy (e-mail: luca.barbisan@polito.it, fabrizio.riente@polito.it)}
\thanks{Marco Levorato is with the Department of Computer Science, Donald Bren School of Information and Computer Sciences, University of California, Irvine, Irvine, CA 92697 USA (e-mail: levorato@uci.edu).}}



\maketitle

\begin{abstract}
Developing algorithms for sound classification, detection, and localization requires large amounts of flexible and realistic audio data, especially when leveraging modern machine learning and beamforming techniques. However, most existing acoustic simulators are tailored for indoor environments and are limited to static sound sources, making them unsuitable for scenarios involving moving sources, moving microphones, or long-distance propagation. This paper presents \textit{DynamicSound} an open-source acoustic simulation framework for generating multichannel audio from one or more sound sources with the possibility to move them continuously in three-dimensional space and recorded by arbitrarily configured microphone arrays. The proposed model explicitly accounts for finite sound propagation delays, Doppler effects, distance-dependent attenuation, air absorption, and first-order reflections from planar surfaces, yielding temporally consistent spatial audio signals. 
Unlike conventional mono or stereo simulators, the proposed system synthesizes audio for an arbitrary number of virtual microphones, accurately reproducing inter-microphone time delays, level differences, and spectral coloration induced by the environment. Comparative evaluations with existing open-source tools demonstrate that the generated signals preserve high spatial fidelity across varying source positions and acoustic conditions. 
By enabling the generation of realistic multichannel audio under controlled and repeatable conditions, the proposed open framework provides a flexible and reproducible tool for the development, training, and evaluation of modern spatial audio and sound-source localization algorithms.
\end{abstract}


\section{Introduction}

\IEEEPARstart{W}{ith} the 
adoption of autonomous vehicles, delivery robots, and intelligent surveillance systems, the ability to analyze the surrounding acoustic environment is becoming increasingly important. In these scenarios, microphones and microphone arrays represent valuable sensing modalities for detecting and interpreting sound events in real time. For instance, single microphones can be used to detect emergency vehicles using neural networks \cite{parineh2023detecting}, or to estimate the speed of an approaching vehicle by exploiting Doppler-induced frequency shifts \cite{bulatovic2022mel, djukanovic2021acoustic}. Microphone arrays, by leveraging their capability to estimate the direction of arrival (DOA) of sound waves, offer even richer information: when placed near a roadway, they can be used to count passing vehicles, estimate their speed, and determine their direction of movement \cite{peral2013using, zu2017vehicle}. Arrays can also be mounted on unmanned aerial vehicles (UAVs) for sound-source localization during search-and-rescue operations \cite{nakadai2017development, hoshiba2017design, go2021acoustic}, or conversely, deployed on the ground to detect and track flying drones \cite{blanchard2020acoustic, manickam2021multi, riabko2024edge}.

These applications highlight the growing need for large, diverse, and realistic acoustic datasets that can support the development and evaluation of robust signal-processing and machine-learning algorithms. In many cases, however, collecting real-world recordings is prohibitively expensive, logistically difficult, or even impossible, especially when recreating rare or hazardous events. Furthermore, microphone arrays can have very different geometries and spatial configurations; being able to simulate these configurations greatly simplifies the study and optimization of new array designs before physically building them.

Several acoustic simulation frameworks are currently available, each addressing specific aspects of sound propagation. Among open-source tools, \textit{pyroomacoustics}\cite{scheibler2018pyroomacoustics} is widely used for simulating room acoustics and microphone arrays in indoor environments. While it provides accurate models for reverberation and reflection in enclosed spaces, it is primarily designed for static scenarios and does not natively support sound sources moving continuously in time. On the other hand, \textit{pyroadacoustics}\cite{damiano2022pyroadacoustics} targets outdoor traffic-noise simulations and introduces support for dynamic scenarios; however, in its current implementation, motion is effectively modeled at the array level, and changes in source dynamics are reflected instantaneously at the receiver, neglecting the finite time of sound propagation. As a result, physically correct delays associated with sound source movement are not achievable and this limitation is more visible for rapidly changing trajectories.

Commercial software packages such as Odeon\cite{naylor1993odeon, odeon}, EASE\cite{ease}, CATT-Acoustic\cite{catt_acoustic}, COMSOL Multiphysics\cite{comsol}, CadnaA\cite{cadnaa}, and SoundPLAN\cite{soundplan} offer high-fidelity acoustic modeling and are widely used in architectural acoustics and noise prediction. Despite their accuracy, these tools are typically proprietary, computationally intensive, and not easily integrated into modern machine-learning workflows. Moreover, they often lack direct support for exporting time-domain multichannel signals of moving sound sources suitable for training neural networks or evaluating array-processing algorithms.

For these reasons, this article presents \textit{DynamicSound}, a  simulation framework developed to provide researchers with a flexible and physically consistent tool for modeling the propagation of sound emitted by one or more moving sources in three-dimensional space. The software, openly available on Github and archived on Zenodo \cite{dynamicsound2025}, is implemented using object-oriented programming principles to ensure modularity and extensibility. The current version supports multiple simultaneous sources and this allows to implement the image-source method to model first-order reflections from planar surfaces. Although the present simulator does not account for occlusion or diffraction effects that become relevant in more complex architectural environments, it offers a realistic and computationally efficient approximation for open-field scenarios with a few reflective surfaces. The main contributions of this work are summarized as follows:
\begin{itemize}
\item We introduce a physically consistent time-domain acoustic simulation model for continuously moving sound sources that explicitly accounts for finite sound propagation delays and Doppler effects, avoiding the instantaneous-update assumptions commonly adopted in existing simulators.
\item We propose a flexible multichannel spatial audio synthesis framework supporting arbitrary microphone array geometries and an unrestricted number of virtual microphones, accurately reproducing motion-dependent inter-microphone time delays, level differences, and spectral coloration.
\item We integrate distance-dependent attenuation, air absorption, and first-order reflections from planar surfaces into a modular and computationally efficient architecture suitable for open and semi-open environments.
\item We provide comparative evaluations against existing open-source acoustic simulators, demonstrating improved spatial fidelity and temporal consistency in dynamic scenarios relevant to beamforming and direction-of-arrival estimation.
\end{itemize}

This article aims to describe the design, the implementation and validation of the proposed acoustic simulator. Section~\ref{sec:intro} introduces the underlying acoustic physics modeled in the framework, including time-of-flight propagation delays, Doppler effects, geometric spreading, and air absorption. Section~\ref{sec:sim_architecture} discusses the software architecture and implementation. Section~\ref{sec:results} presents comparative evaluations against other open-source Python-based acoustic simulators across different scenarios. Finally, Section~\ref{sec:conclusions} concludes the paper and outlines potential future extensions.

\section{Acoustic Propagation Modeling}
\label{sec:intro}

The framework incorporates four key phenomena that govern sound propagation in open-field environments: time-of-flight propagation delays, Doppler effects, geometric spreading, and air absorption. These components ensure that the generated multichannel audio signals accurately reflect the spatial and spectral characteristics produced by a real sound source moving in three-dimensional space.

\subsection{Time-of-Flight Propagation Delay}
When a sound source emits an acoustic signal, the pressure waves propagate through the medium at a finite speed, typically approximated by the speed of sound in air computed using Equation \ref{eq:sound_speed} that depends on the temperature \(T_K\), expressed in Kelvin.

\begin{equation}
    c_{air} = 343.2 \times \sqrt{\frac{T_K}{293.15}}
    \label{eq:sound_speed}
\end{equation}

For a stationary source located at position $\vect{p}_s$ and a receiver at position $\vect{p}_r$, the propagation delay $\Delta(t)$ is computed as:

\begin{equation}
    \Delta t = \frac{\|\vect{p}_s - \vect{p}_r\|}{c},
\end{equation}

where $c$ denotes the speed of sound. This time-varying delay allows the simulator to reproduce inter-microphone time differences, which are critical for beamforming and direction-of-arrival estimation.

In a generic scenario where both the sound source and the microphone receiver may move with constant linear velocities, the time-of-flight relationship becomes time-dependent. To determine the emission time $t_e$ corresponding to a given observation time $t_r$ at the receiver position $\vect{p}_r(t_r)$, we use the propagation-delay reported in equation \ref{eq:travel_time} that accounts for the distance between the moving source and receiver:
\begin{equation}
    ||\vect{p}_r(t_r) - \vect{p}_s(t_e)|| = c \cdot (t_r - t_e).
    \label{eq:travel_time}
\end{equation}
Valid solutions exist only for $t_e \le t_r$, where $\vect{p}_s(t_e)$ and $\vect{p}_r(t_r)$ denote the positions of the source and microphone at times $t_e$ and $t_r$, respectively, and $c$ is the speed of sound. This formulation enables the simulator to correctly model time-varying delays resulting from motion of either or both entities.

\subsection{Doppler frequency shift}
Relative motion between the source and microphone introduces a compression or dilation of the acoustic waveform, perceived as a frequency shift. The simulator implements the classical Doppler model, where the instantaneous frequency $f'$ received at the microphone is given by:

\begin{equation}
    f' = \frac{(c + v_r)}{(c - v_s)} f,
\end{equation}
where $f$ denotes the emitted frequency, $c$ is the speed of sound in air, $v_s$ is the source velocity
(positive when moving toward the receiver), and $v_r$ is the receiver velocity (positive when moving toward the source). This effect results in a perceptible pitch increase during approach and a decrease during recession, influencing both spectral and temporal signal characteristics.

\subsection{Spreading attenuation}
As sound propagates, its intensity decreases due to the expansion of the acoustic wavefront. Assuming spherical spreading, the geometric amplitude attenuation factor $A_g(r)$ for a propagation distance $r$ is modeled as:
\begin{equation}
    A_g(r) = \frac{\vect{r}_0}{\vect{r}}s_0(t),
\end{equation}
where $r$ is the actual distance and $r_0$ is the distance of the recording of the $s_0(t)$ signal, resulting in a distance-dependent reduction in signal magnitude. This phenomenon is crucial for recreating realistic amplitude differences among microphones in an array and for matching the behavior of real outdoor sound propagation.

\subsection{Air absorption}
\begin{figure}[!ht]
    \centering
    \includegraphics[width=\linewidth]{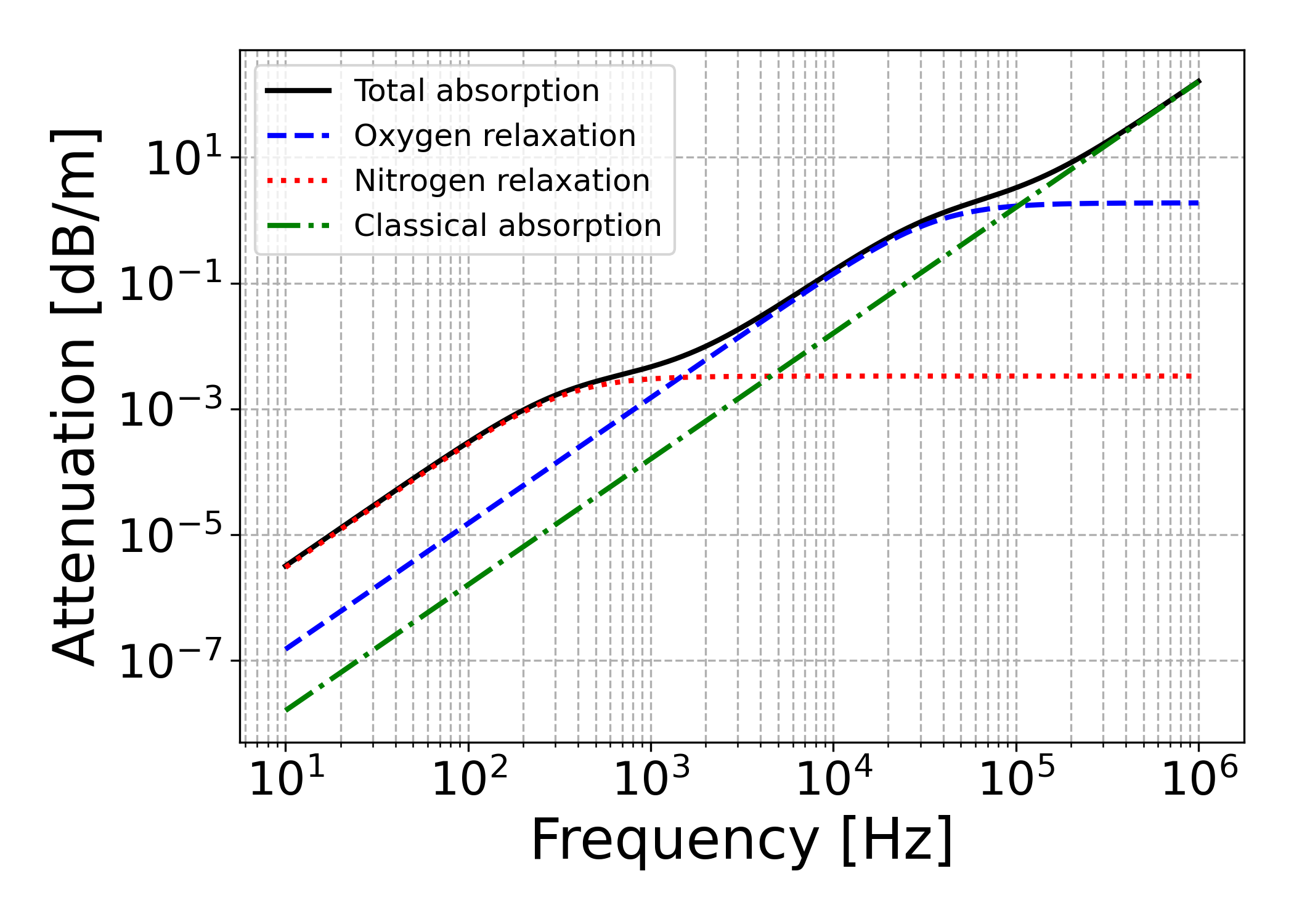}
    \caption{Air absorption at 20 °C, 1 atm, and 50\% relative humidity, showing the total attenuation curve together with the individual oxygen and nitrogen relaxation contributions, computed according to ISO 9613-1 \cite{international1993acoustics}.}
    \label{fig:air_absorption}
\end{figure}

In addition to geometrical spreading, high-frequency components of sound are attenuated due to viscous and thermal losses in air. The attenuation coefficient $\alpha (f)$ depends on frequency, temperature, humidity, and atmospheric pressure, and is computed following the ISO 9613-1 standard \cite{international1993acoustics}. The decrease in sound pressure level caused by air absorption over a propagation distance $d$ can be expressed as:
\begin{equation}
    A_{\mathrm{air}}(f, d) = e^{-\alpha(f)\, d},
\end{equation}
where $\alpha(f)$ is the absorption coefficient, dependent on humidity, temperature, and atmospheric pressure. This effect ensures that long-distance propagation is reproduced with realistic spectral coloration. This frequency-dependent attenuation results in a low-pass filtering effect, especially significant for long distances or high frequencies as shown in Fig.~\ref{fig:air_absorption}.

According to the ISO 9613-1 standard \cite{international1993acoustics}, the absorption coefficients $\alpha_0$ are defined by the expression shown in equation \ref{eq:air_absorption_coeff}.

\scalebox{0.875}{
\parbox{\linewidth}{
\begin{multline}
\alpha(f,T,p_a,h) = 8.686 f^2 \Bigg(
\left[ 1.84 \times 10^{-11} \left( \frac{p_{s0}}{p_a} \right) \left( \frac{T_0}{T} \right)^{1/2} \right]\\ +
\left( \frac{T}{T_0} \right)^{-5/2} \times
\Bigg\{ 0.01275 \left[ \exp \left( \frac{-2239.1}{T} \right) \right] \left[ f_{rO} + \left( \frac{f^2}{f_{rO}} \right) \right]^{-1}\\ +
0.1068 \left[ \exp \left( \frac{-3352.0}{T} \right) \right] \left[ f_{rN} + \left( \frac{f^2}{f_{rN}} \right) \right]^{-1} \Bigg\} \Bigg)
\label{eq:air_absorption_coeff}
\end{multline}
}}

 In this formulation, $f$ denotes the source signal frequency, $T$ is the air temperature, $T_0 = 293.15 \mathrm{K}$ is the reference temperature, $p_a$ represents the atmospheric pressure, $p_{s0} = 101.325 \mathrm{kPa}$ is the international standard atmospheric pressure at mean sea level, $f_{r\mathrm{O}}$ is the oxygen relaxation frequency computed with equation~\ref{eq:oxygen_relaxation} and $f_{r\mathrm{N}}$ is the nitrogen relaxation frequency computed with equation~\ref{eq:nitrogen_relaxation}. Finally, the molar concentration of water vapor $h$ can be computed in terms of relative humidity $h_{rel}$ using the equation~\ref{eq:molar_concentration} that depends on the saturation pressure $p_{sat}$ expressed by the equation~\ref{eq:saturation_pressure} where $T_{01}=273.16 \mathrm{K}$ is the triple-point isotherm temperature.

\begin{equation}
    f_{rO} = \frac{p_a}{p_r} \left( 24 + 4.04 \cdot 10^4 \cdot h \cdot \frac{0.02 + h}{0.391 + h} \right)
    \label{eq:oxygen_relaxation}
\end{equation}

\begin{equation}
    f_{rN} = \frac{p_a}{p_r} \left( \frac{T}{T_0} \right) ^{-1/2} \left( 9 + 280 \cdot h \cdot e^{ -4.170 \left[ \left( \frac{T}{T_0} \right) ^{-1/3} -1 \right] } \right)
    \label{eq:nitrogen_relaxation}
\end{equation}

\begin{equation}
    h = h_{rel} \cdot \frac{p_{sat}}{p_s}
    \label{eq:molar_concentration}
\end{equation}

\begin{equation}
    p_{sat} = p_{s0} \cdot 10 ^ { \left[ -6.8346 \left( \frac{T_{01}}{T} \right)^{1.261} + 4.6151 \right] }
    \label{eq:saturation_pressure}
\end{equation}


\subsection{Reflections}
Reflections from surfaces such as the ground, buildings, or vegetation alter the spatial and spectral characteristics of the received sound through interference and diffusion. Each reflection is characterized by its reflection coefficient $R(f)$, which depends on the material impedance and incidence angle.

\begin{figure}[!htbp]
    \centering
    \includegraphics[width=0.85\linewidth]{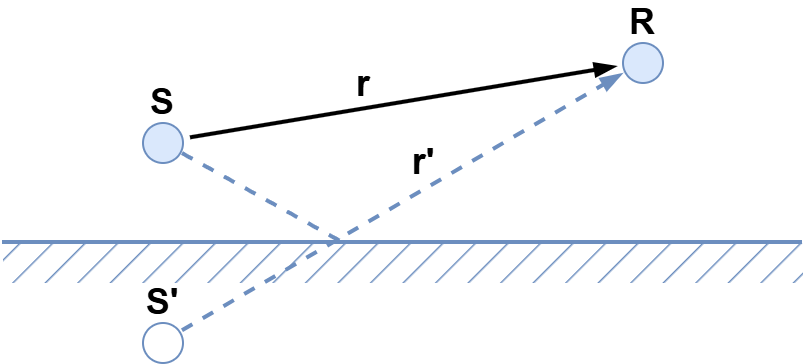}
    \caption{Illustration of the image-source method used to model reflections, showing the real source \textbf{S}, its mirrored image source \textbf{S'}, and the resulting direct and reflected propagation paths to the receiver \textbf{R}.}
    \label{fig:reflection_img}
\end{figure}

Specular reflections produce comb-filtering effects due to path-length differences. In this version of the simulator, reflections can be modeled using image-source methods creating a copy of the source path but with a specular position with respect to the reflection plane. Fig.~\ref{fig:reflection_img} illustrates an example in which the real source \textbf{S} emits sound that reaches the receiver \textbf{R} along the direct path $\vect{r}$, while the reflected sound is modeled by a virtual source \textbf{S'} obtained by mirroring the trajectory of \textbf{S} with respect to the horizontal plane. This reflected position requires a longer path $\vect{r'}$ to reach the receiver, resulting in increased propagation delay and attenuation.

\begin{figure*}[!ht]
    \centering
    \includegraphics[width=\linewidth]{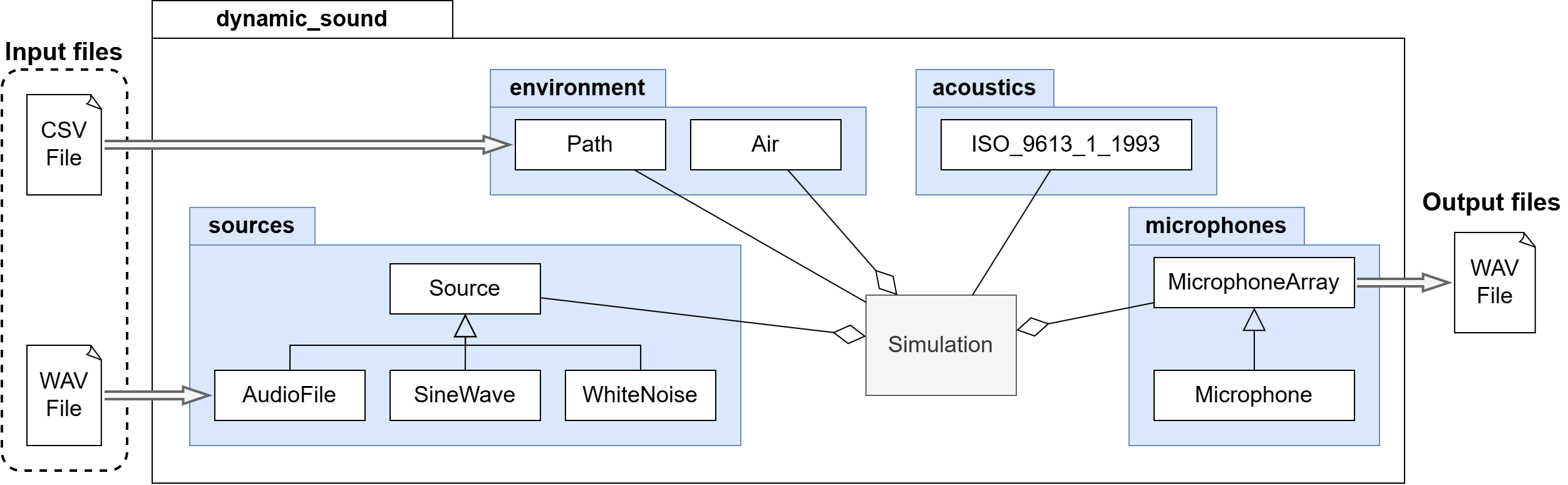}
    \caption{Class diagram of the \textit{DynamicSound} simulator architecture, illustrating the overall software structure and the interactions between the main classes. The diagram highlights the separation into distinct packages: \textit{microphones}, \textit{sources}, \textit{environment}, and \textit{acoustics}.}
    \label{fig:sim_diagram}
\end{figure*}

\section{Simulator Architecture}
\label{sec:sim_architecture}

The overall structure of the simulator is illustrated in Fig.~\ref{fig:sim_diagram}. The architecture is organized into four packages: microphones, sources, environment, and acoustics packages. The \textit{microphones} package allows the user to instantiate either a single virtual microphone or a microphone array that can be moved and rotated as a unique object during the simulations. At the end of the simulation, every instance of the \textit{Microphone} and \textit{MicrophoneArray} classes generates its own audio file where for each microphone a separate audio channel is created. The sampling rate can be freely specified, enabling simulations at any desired frequency resolution. 

The \textit{sources} package provides different classes for generating acoustic signals. In the current version, three source types are available: a white-noise generator, a sinusoidal signal generator, and an audio-file class that allows loading arbitrary sound files from disk. These sources can be placed statically or moved along predefined trajectories.

The \textit{acoustics} package encapsulates the core physical models and standardized formulations used throughout the simulator, including implementations of international acoustic standards, propagation laws, and a collection of utility functions for computing time delays, attenuation, Doppler effects, and frequency-dependent air absorption.

Finally, the \textit{environment} module contains all information related to the propagation medium and spatial configuration. This includes the source and microphone paths as well as atmospheric parameters such as temperature, pressure, and relative humidity, which govern the air-absorption model and other propagation effects. Together, these modules form a flexible and extensible framework for simulating realistic acoustic scenes in both static and dynamic conditions.

\subsection{Computation of the emission time $t_e$}

The emission time of the acoustic signal was obtained by solving the travel-time equation \ref{eq:travel_time}. Expanding and squaring both sides leads to a quadratic equation in $t_e$:
\begin{equation}
    \underbrace{(||v_s||^2-c^2)}_{\text{a}}t_e^2 + \underbrace{2(c^2t_r-r_0\cdot v_s)}_{\text{b}}t_e + \underbrace{(||r_0||^2-c^2t_r^2)}_{\text{c}} = 0,
\label{eq:quadratic_formula}
\end{equation}
where $\mathrm{r_0}$ is defined as $r_0 \equiv p_r - p_{s,0}$, the initial relative distance between source and receiver. The discriminant $\Delta = b^2 - 4ac$ determines the number of real solutions.

\begin{equation}
    t_{e1,2} = \frac{-b \pm \sqrt{\Delta}}{2a}
\end{equation}

From this quadratic equation only the physically admissible emission time will be considered, the real solutions that satisfy the $t_e \le t_r$ ensuring causality of the system.

\subsection{Source position at emission time.}

Once the emission time $t_e$ is computed, the corresponding position of the moving source is obtained directly from its kinematic model. Assuming linear motion with constant velocity, the source position at the emission is:
\begin{equation}
    \mathbf{p}_s(t_e) = \mathbf{p}_{s0} + \mathbf{v}_s\,t_e,
\end{equation}
where $\mathbf{p}_{s0}$ is the known position of the source at the reference time $t = 0$ and $\mathbf{v}_s$ is its velocity vector. This formulation ensures spatial and temporal consistency between the estimated emission event and the received signal. The computed $\mathbf{p}_s(t_e)$ thus represents the physical location of the source at the instant the acoustic wavefront that reached the receiver at time $t_r$ was emitted.

\subsection{Time interpolation}
In a moving-source localization scenario, the time delay at the receiver plays a crucial role. To accurately simulate this behavior, the delay must vary smoothly in response to the changing position of the source. When the signal is described by a mathematical function, such as a sinusoidal waveform, the corresponding sample value can be computed for any $t \in \mathbf{R}$. However, when the sound source is represented by a discretely sampled signal as in common audio formats, such as the waveform audio (WAV) files, each sample is available only at integer indices. In this case, interpolation becomes essential to obtain values at intermediate positions. In the current implementation, linear interpolation was found to provide sufficient quality when the source signal has a sufficiently high sampling rate. Otherwise, the signal can be resampled during its creation by specifying a new sampling rate and a resampling method provided by the \textit{Librosa} library \cite{mcfee2015librosa}.


\subsection{Motion paths}
\begin{figure}[!ht]
    \centering
    \includegraphics[width=\linewidth]{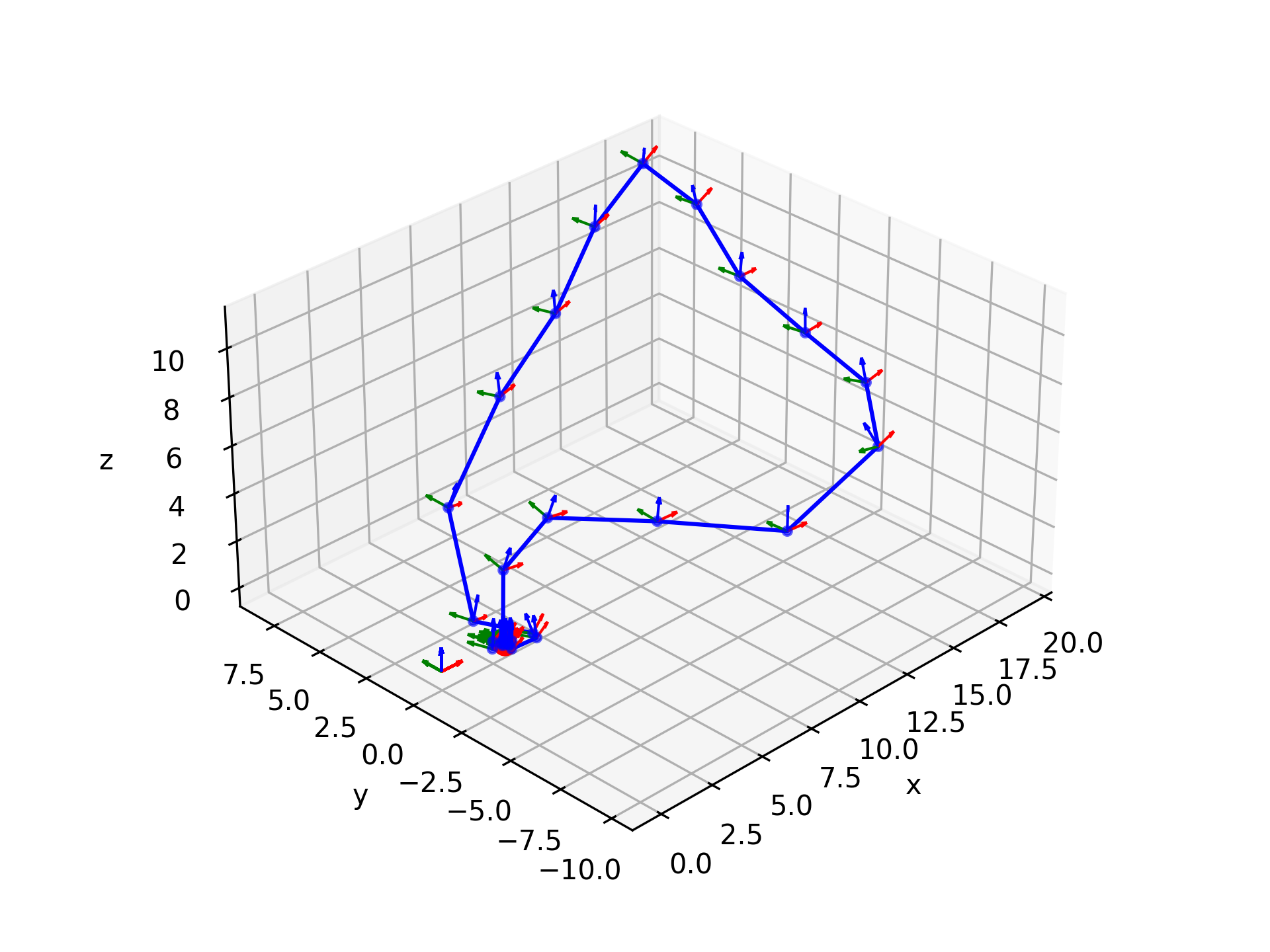}
    \caption{Drone trajectory generated using \textit{AirSim} \cite{shah2017airsim}, sampled at 1-second intervals throughout the physical simulation.}
    \label{fig:drone_path}
\end{figure}

\begin{figure}[!ht]
    \centering
    \includegraphics[width=\linewidth]{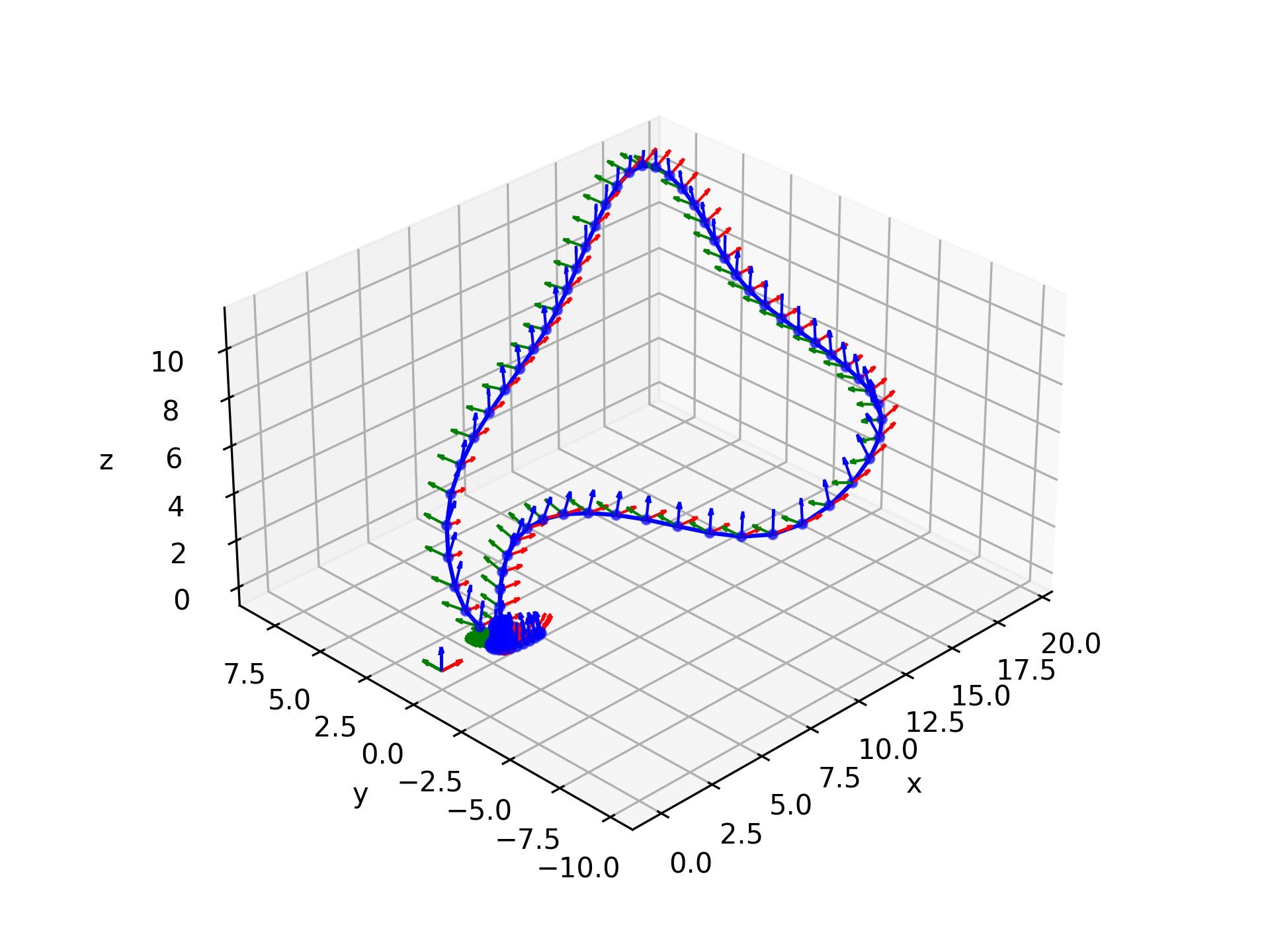}
    \caption{Interpolated drone trajectory produced using the \textit{DynamicSound} simulator, providing a smooth path reconstruction from the original 1-second \textit{AirSim} samples.}
    \label{fig:drone_path_interpolated}
\end{figure}

The paths describing the motion of the objects are defined as lists of positions and rotations at specific time instants. These paths can also be loaded from a CSV file that can be generated by another tool like the path in Fig.~\ref{fig:drone_path} which is generated with the AirSim physics simulator and represents the trajectory of a quadcopter. Between each pair of points, the motion is assumed to be linear with constant velocity, and the orientation is also interpolated linearly. To achieve better audio results, especially when the direction or speed changes significantly, it is possible to generate intermediate positions by interpolating the path data, resulting in a smoother trajectory, as illustrated in Fig.~\ref{fig:drone_path_interpolated}.

\subsection{Air absorption filter}

The air absorption coefficients are calculated according to Equation~\ref{eq:air_absorption_coeff}. A discrete set of coefficients is distributed over the frequency range from $0$ to $\text{sample\_rate}/2$. Increasing the number of frequency samples enhances the accuracy of the model by providing a finer spectral resolution.

After computation, the absorption coefficients are scaled by the propagation distance and subsequently converted from decibel to linear magnitude. The resulting linear coefficients are then employed to design the corresponding finite impulse response (FIR) filter, which models the frequency-dependent attenuation due to air absorption.

\begin{figure*}[!htp]
    \centering

    \begin{minipage}{0.32\textwidth}
        \centering
        \begin{overpic}[width=\linewidth]{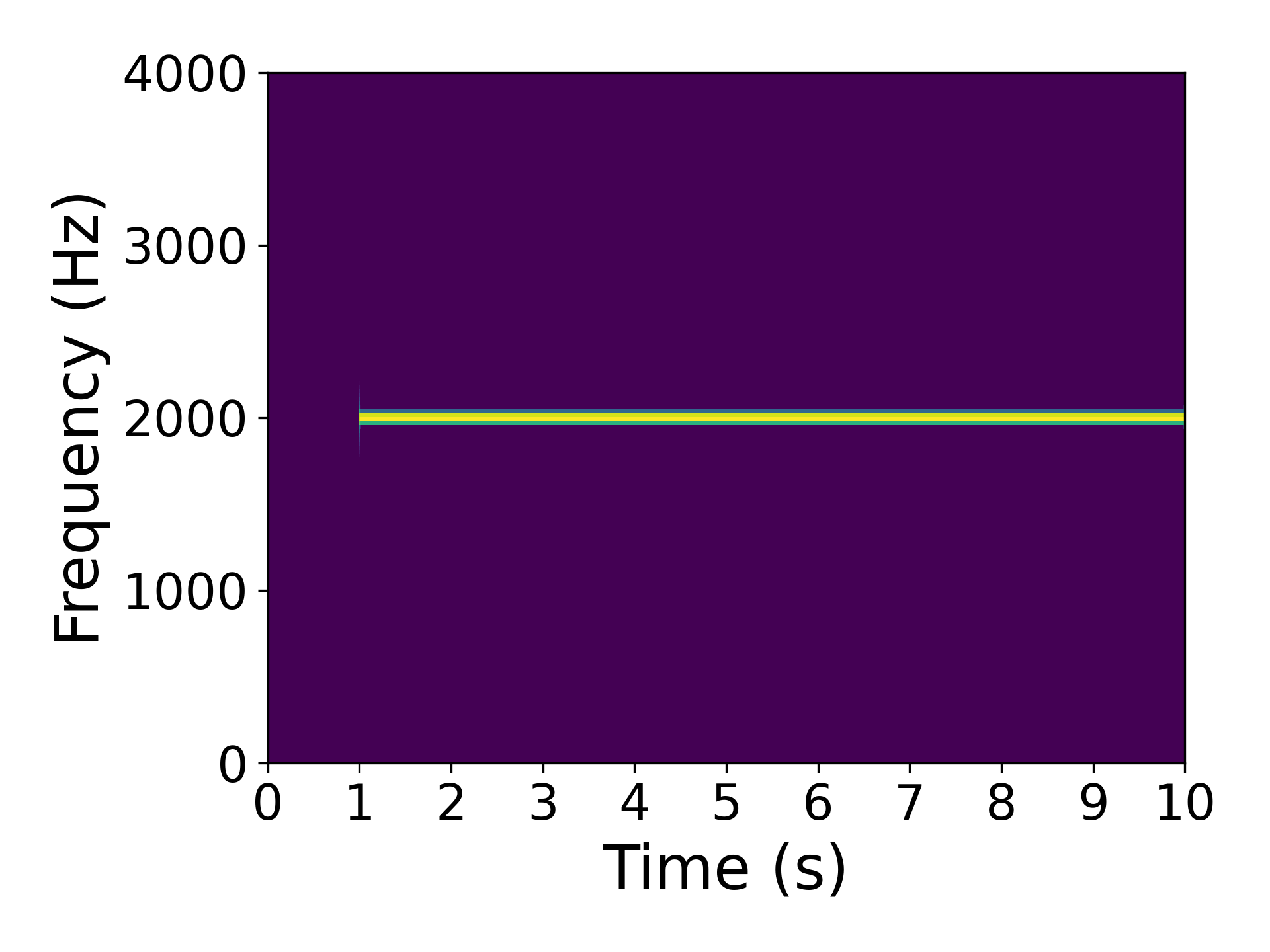}
            \put(-5,65){\colorbox{white}{\textbf{a)}}}
        \end{overpic}
    \end{minipage}\hfill
    \begin{minipage}{0.32\textwidth}
        \centering
        \begin{overpic}[width=\linewidth]{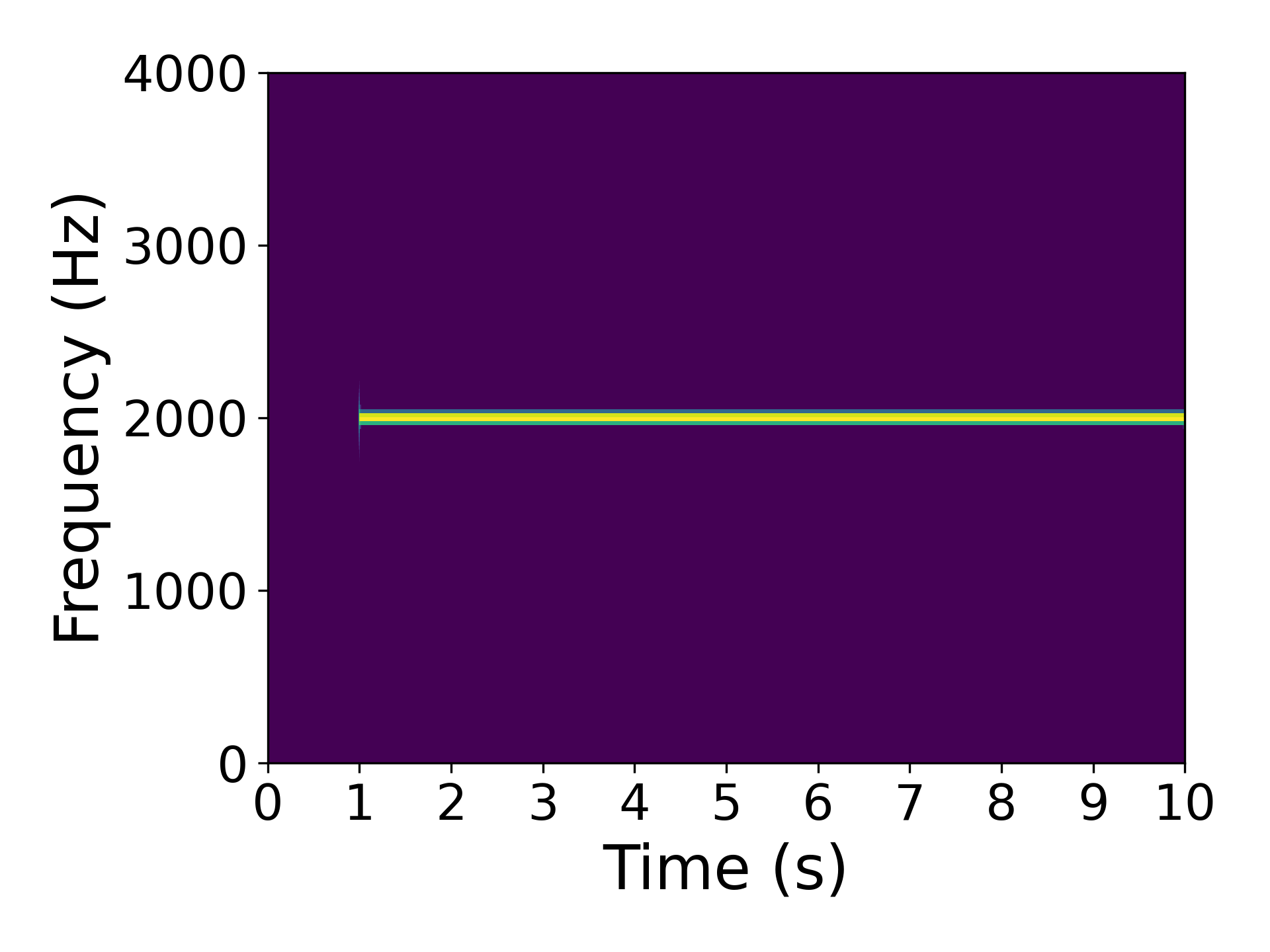}
            \put(-5,65){\colorbox{white}{\textbf{b)}}}
        \end{overpic}
    \end{minipage}\hfill
    \begin{minipage}{0.32\textwidth}
        \centering
        \begin{overpic}[width=\linewidth]{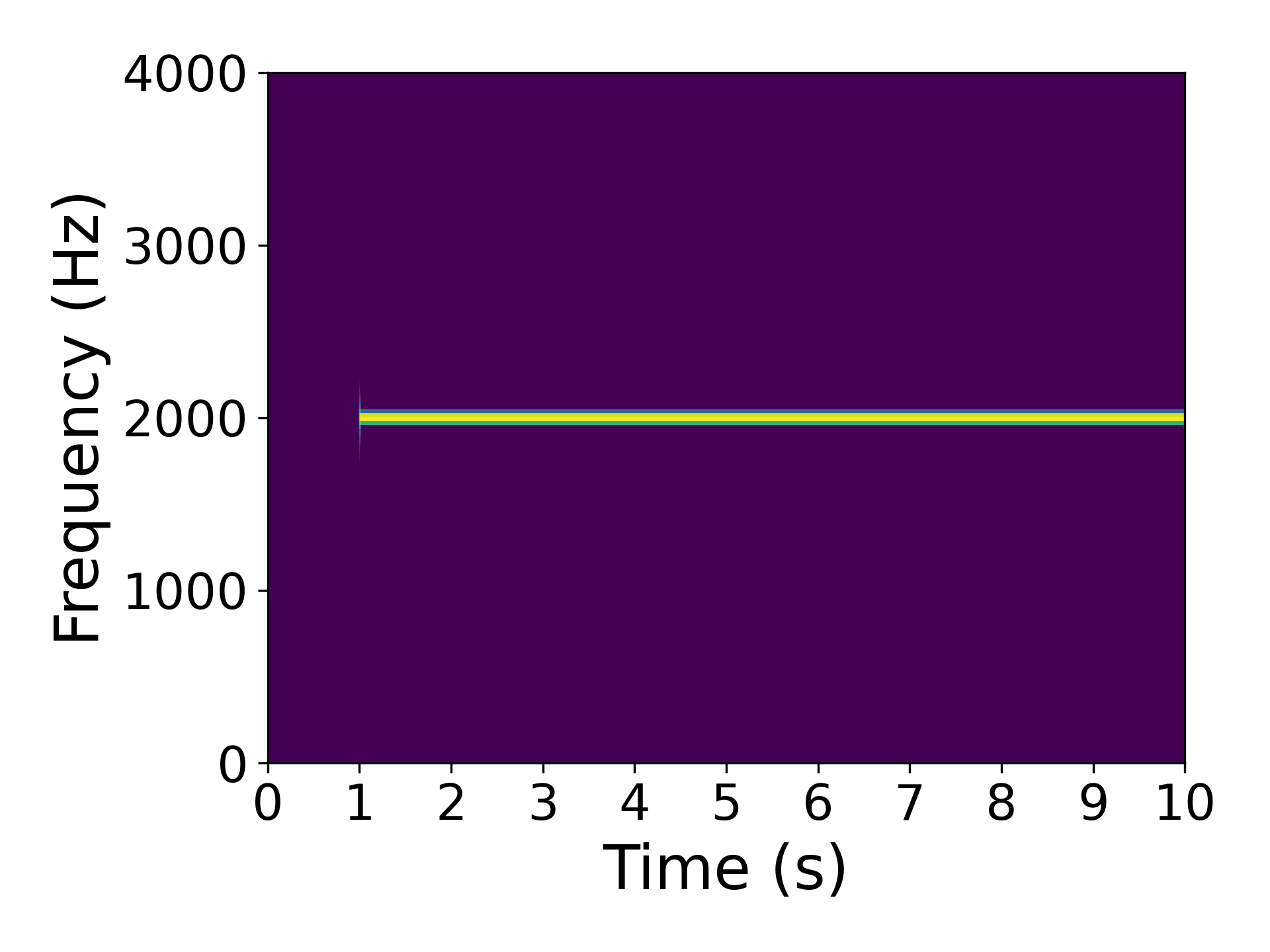}
            \put(-5,65){\colorbox{white}{\textbf{c)}}}
        \end{overpic}
    \end{minipage}

    \vspace{0em}

    \begin{minipage}{0.32\textwidth}
        \centering
        \begin{overpic}[width=\linewidth]{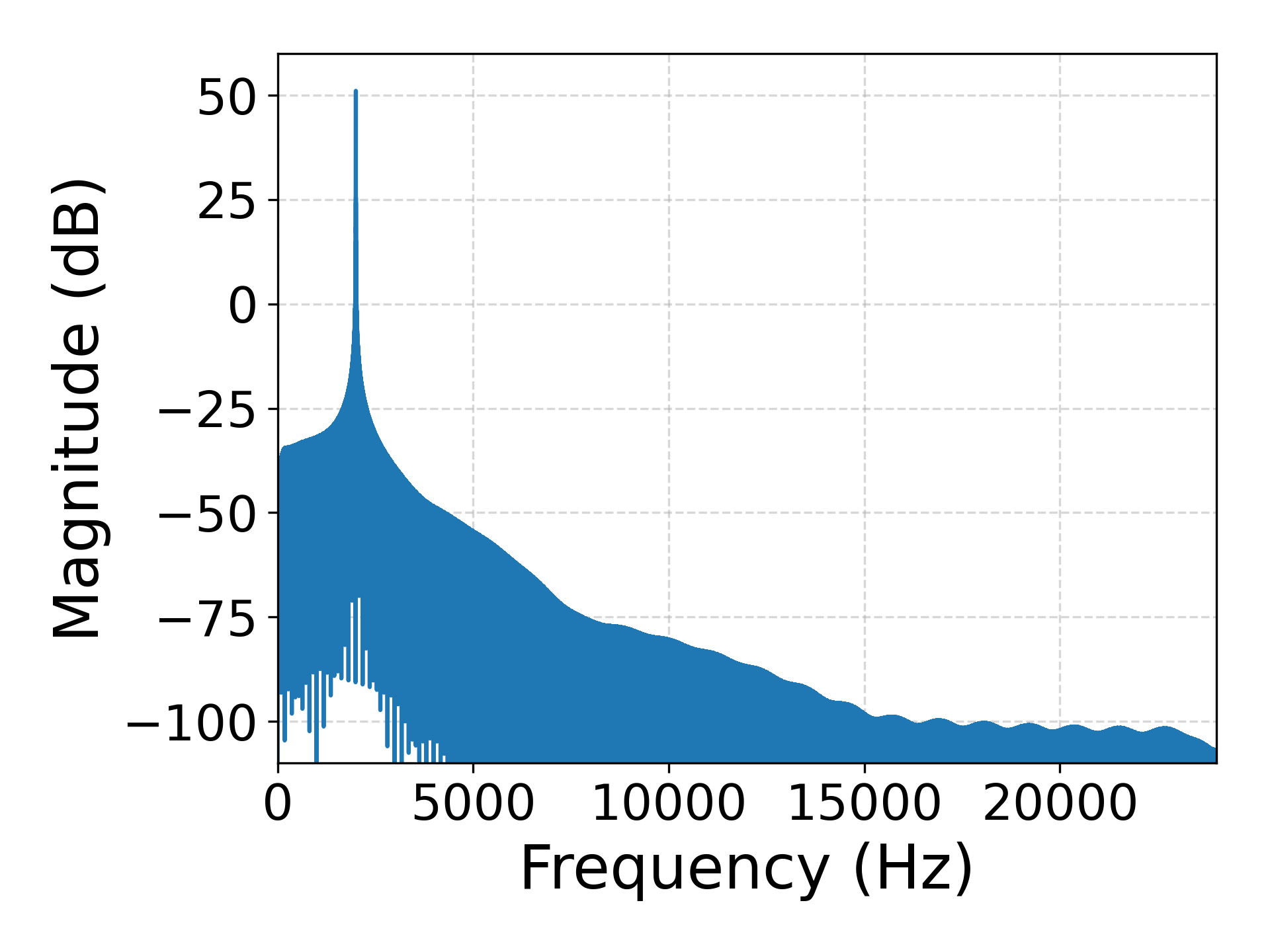}
            \put(-5,65){\colorbox{white}{\textbf{d)}}}
        \end{overpic}
    \end{minipage}\hfill
    \begin{minipage}{0.32\textwidth}
        \centering
        \begin{overpic}[width=\linewidth]{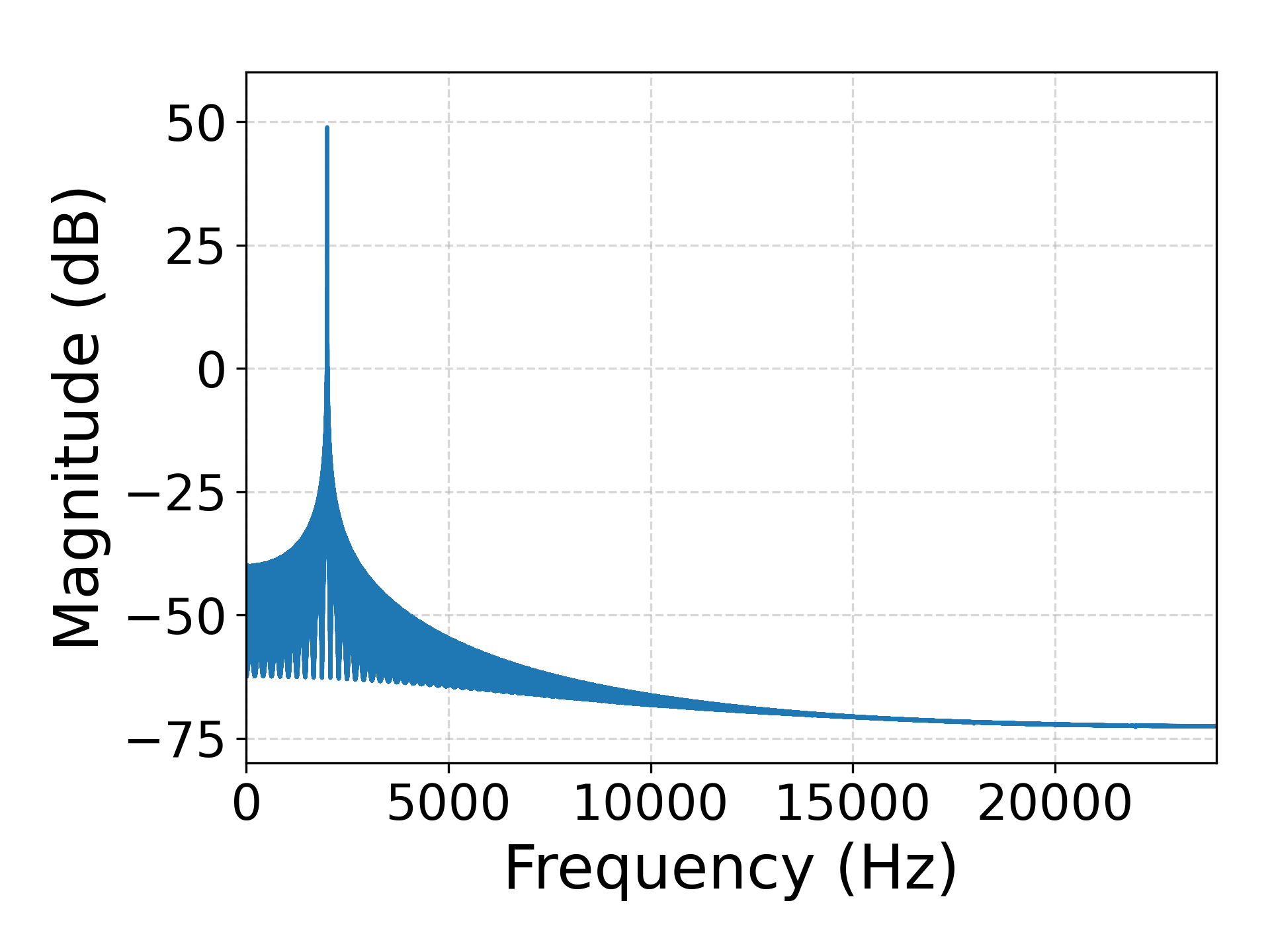}
            \put(-5,65){\colorbox{white}{\textbf{e)}}}
        \end{overpic}
    \end{minipage}\hfill
    \begin{minipage}{0.32\textwidth}
        \centering
        \begin{overpic}[width=\linewidth]{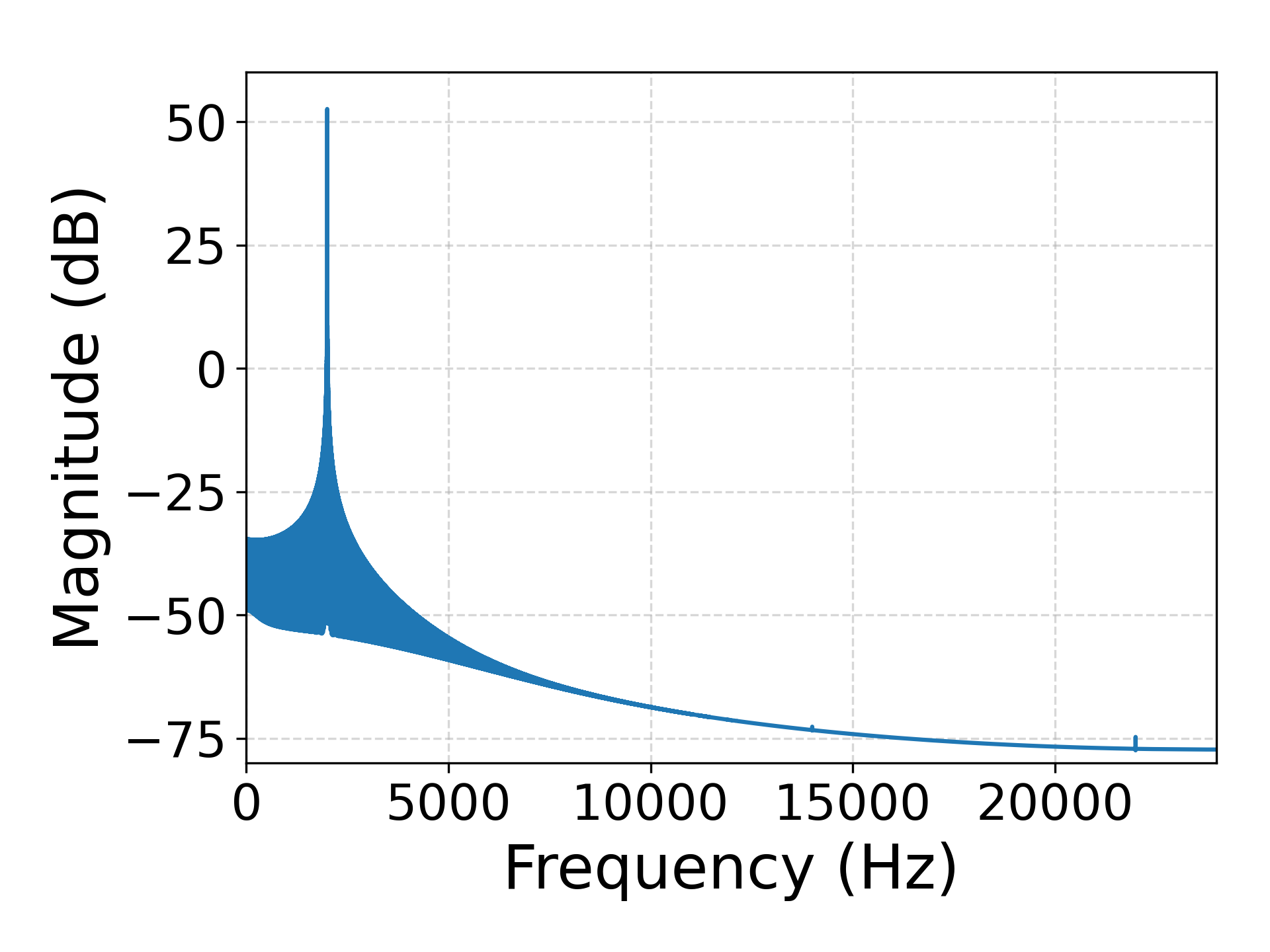}
            \put(-5,65){\colorbox{white}{\textbf{f)}}}
        \end{overpic}
    \end{minipage}
    
    \caption{Sinusoidal signal sound source at \(434\mathrm{m}\) from the microphone. a) spectrogram and d) FFT of the audio generated with \textit{pyroomacoustics}. b) spectrogram and e) FFT of the audio generated with \textit{pyroadacoustics}. c) spectrogram and f) FFT of the audio generated with \textit{DynamicSound} simulator.}
    \label{fig:static_sin}
\end{figure*}

    

\section{Performance and Results}
\label{sec:results}

This section presents a series of experiments designed to evaluate the accuracy and behavior of the proposed simulator in comparison with two open-source Python frameworks, \textit{pyroomacoustics} \cite{scheibler2018pyroomacoustics} and \textit{pyroadacoustics} \cite{damiano2022pyroadacoustics}, capable of modeling acoustic propagation. Unless otherwise specified, all simulations are conducted under the same atmospheric conditions: a temperature of $20^{\circ}$C, relative humidity of $50\%$, and an ambient pressure of $1$~atm. Although all three tools support customizable environmental parameters, these fixed conditions allow for a consistent comparison across experiments. To ensure reproducibility, the results and figures shown here can be generated using the Jupyter notebooks included in the \texttt{examples} folder of the published repository \cite{dynamicsound2025}.

\subsection{Propagation of a Sinusoidal Source}

The first experiment consists of a single-frequency sinusoidal source of $2\mathrm{\,kHz}$ positioned at a distance of 343\,m from the microphone. As shown in Fig.~\ref{fig:static_sin}(a, b, c), all three simulators correctly exhibit a $1\mathrm{\,s}$ delay between emission and reception, consistent with the time-of-flight expected at that distance. The corresponding frequency-domain analysis Fig.~\ref{fig:static_sin}(d, e, f) confirms that the dominant spectral component aligns with the frequency of the emitted sinusoid, demonstrating that each simulator preserves the primary tonal content of the signal during long-range propagation.

\subsection{Propagation of Broadband Noise}

\begin{figure*}[!htp]
    \centering

    \begin{minipage}{0.32\textwidth}
        \centering
        \begin{overpic}[width=\linewidth]{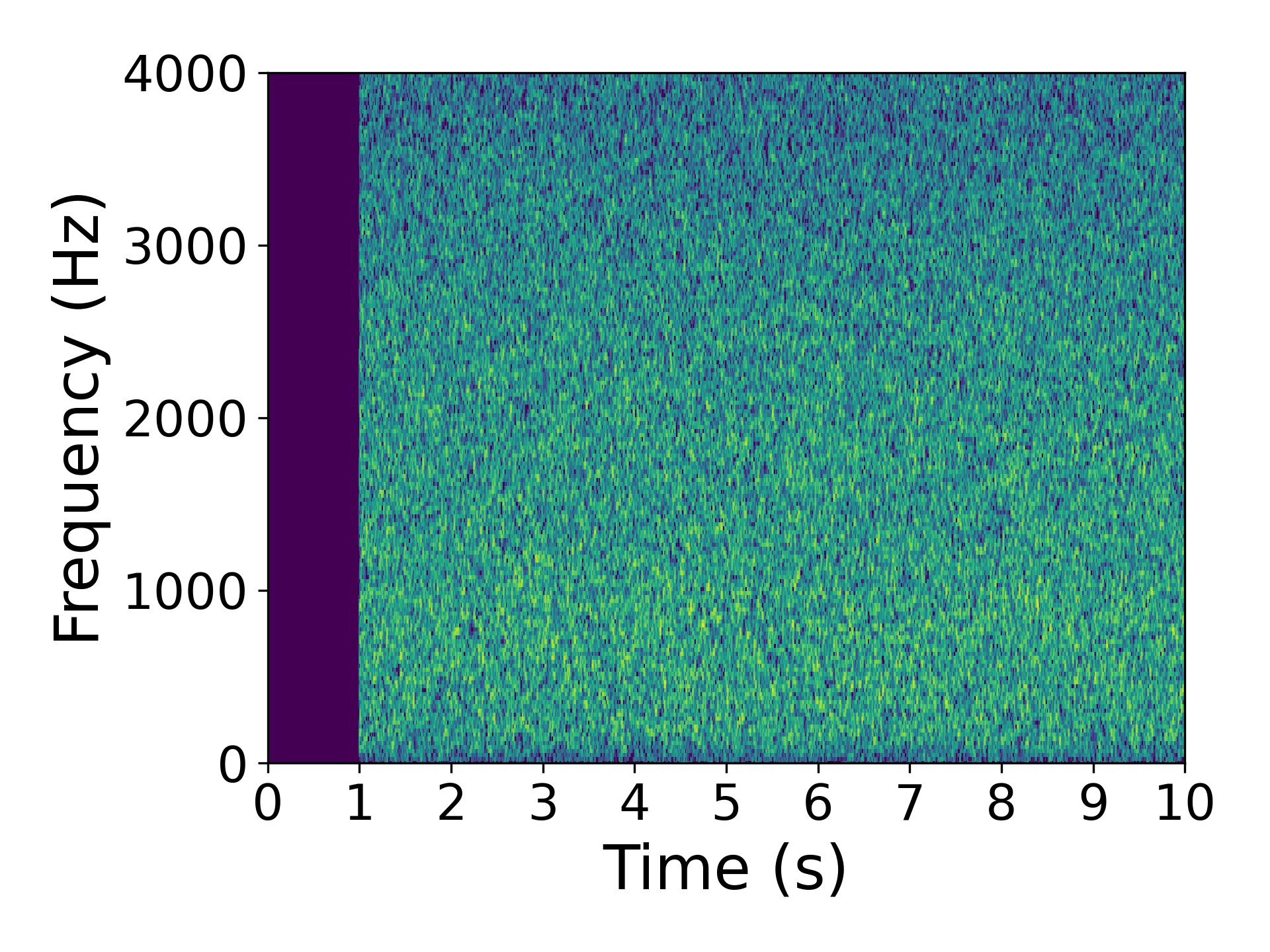}
            \put(-5,65){\colorbox{white}{\textbf{a)}}}
        \end{overpic}
    \end{minipage}\hfill
    \begin{minipage}{0.32\textwidth}
        \centering
        \begin{overpic}[width=\linewidth]{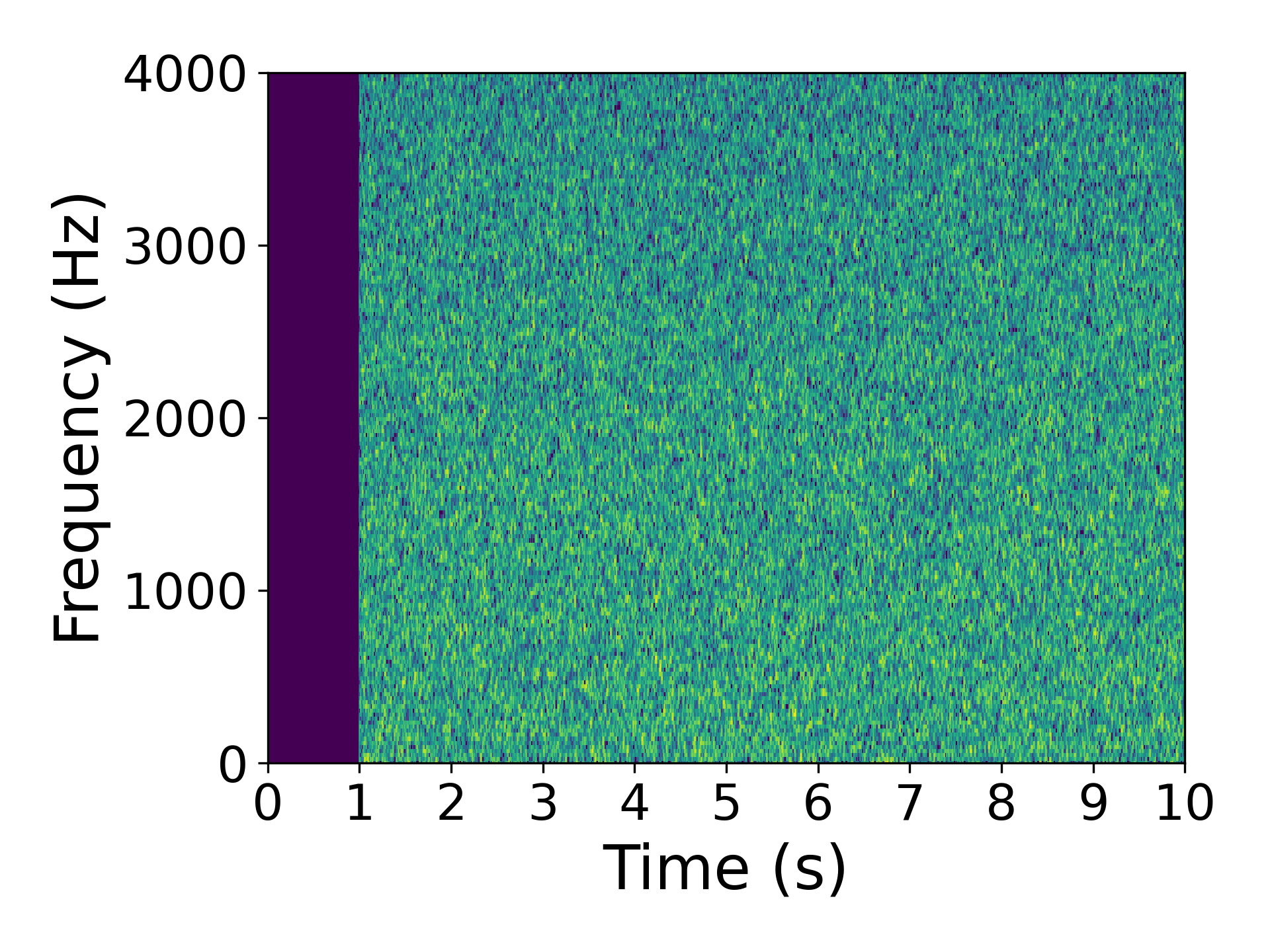}
            \put(-5,65){\colorbox{white}{\textbf{b)}}}
        \end{overpic}
    \end{minipage}\hfill
    \begin{minipage}{0.32\textwidth}
        \centering
        \begin{overpic}[width=\linewidth]{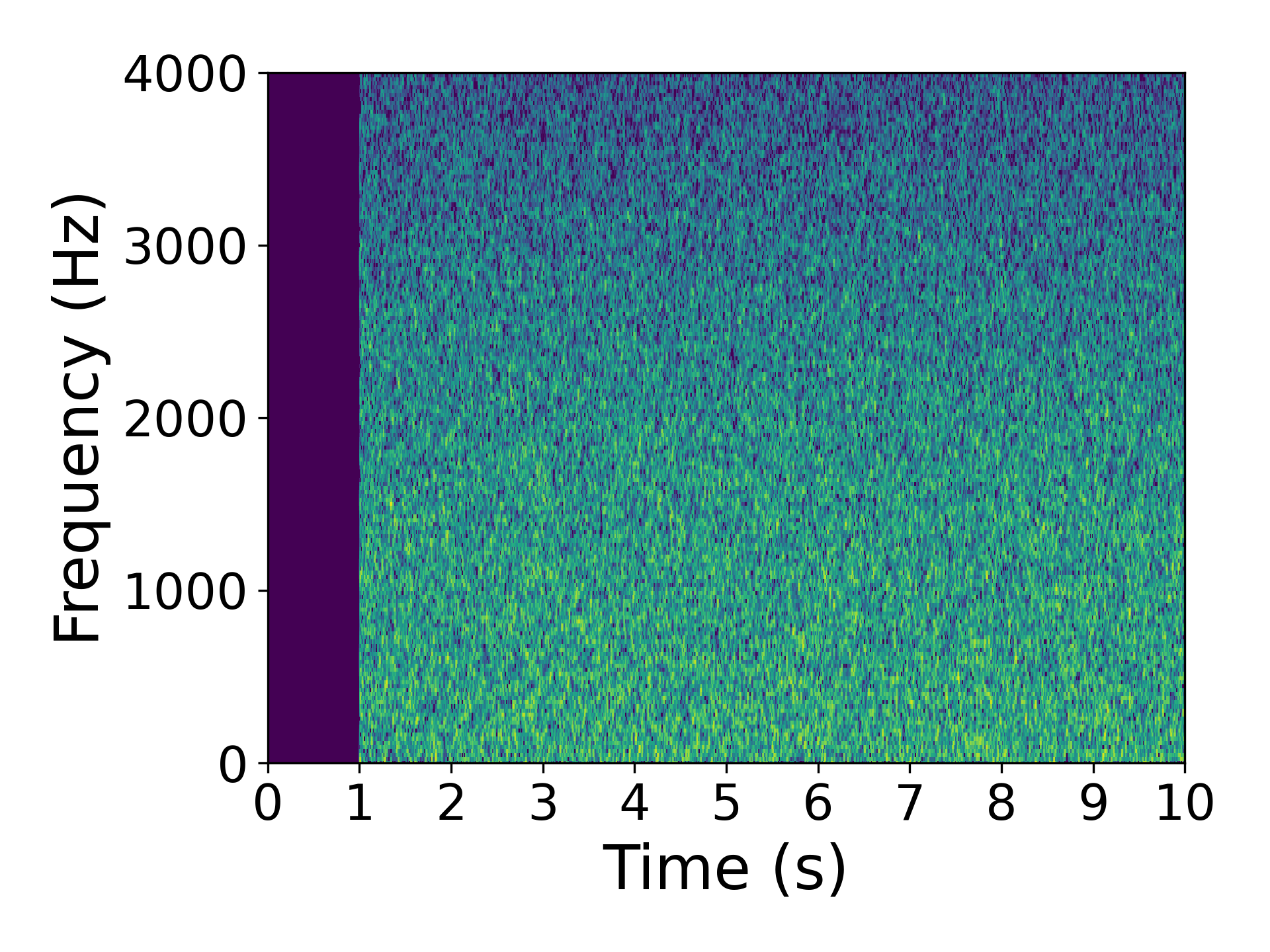}
            \put(-5,65){\colorbox{white}{\textbf{c)}}}
        \end{overpic}
    \end{minipage}

    \vspace{0em}

    \begin{minipage}{0.32\textwidth}
        \centering
        \begin{overpic}[width=\linewidth]{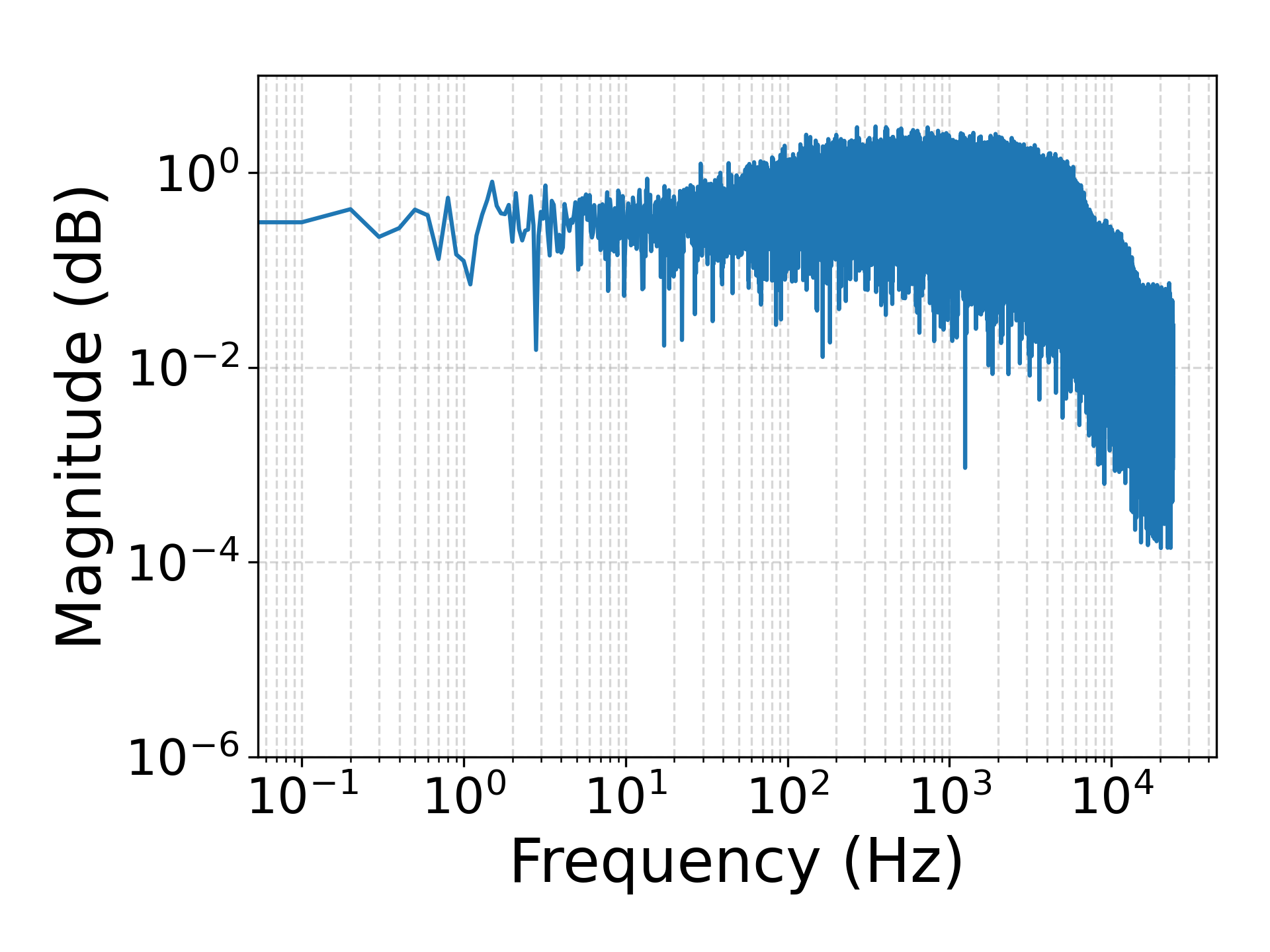}
            \put(-5,65){\colorbox{white}{\textbf{d)}}}
        \end{overpic}
    \end{minipage}\hfill
    \begin{minipage}{0.32\textwidth}
        \centering
        \begin{overpic}[width=\linewidth]{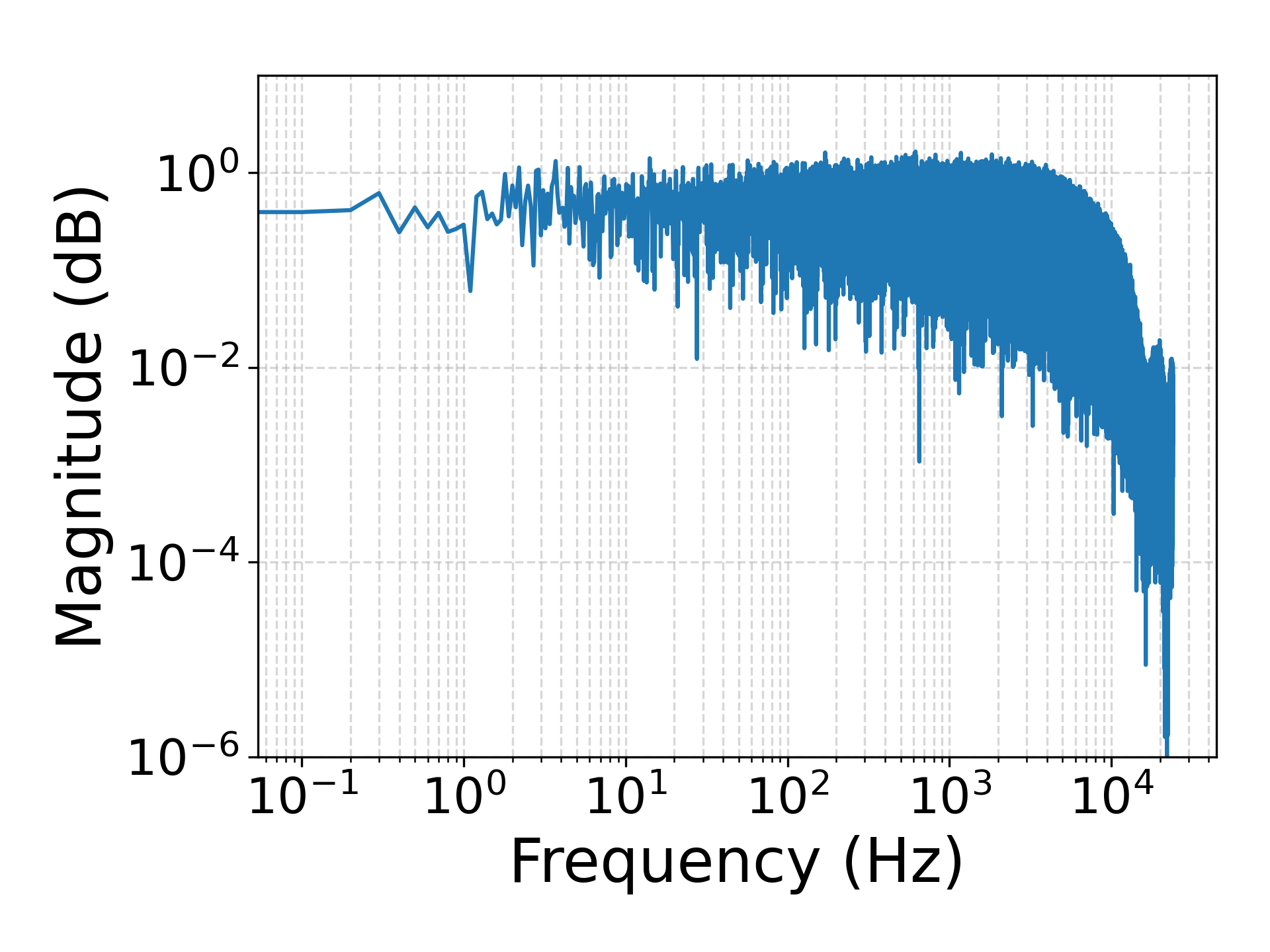}
            \put(-5,65){\colorbox{white}{\textbf{e)}}}
        \end{overpic}
    \end{minipage}\hfill
    \begin{minipage}{0.32\textwidth}
        \centering
        \begin{overpic}[width=\linewidth]{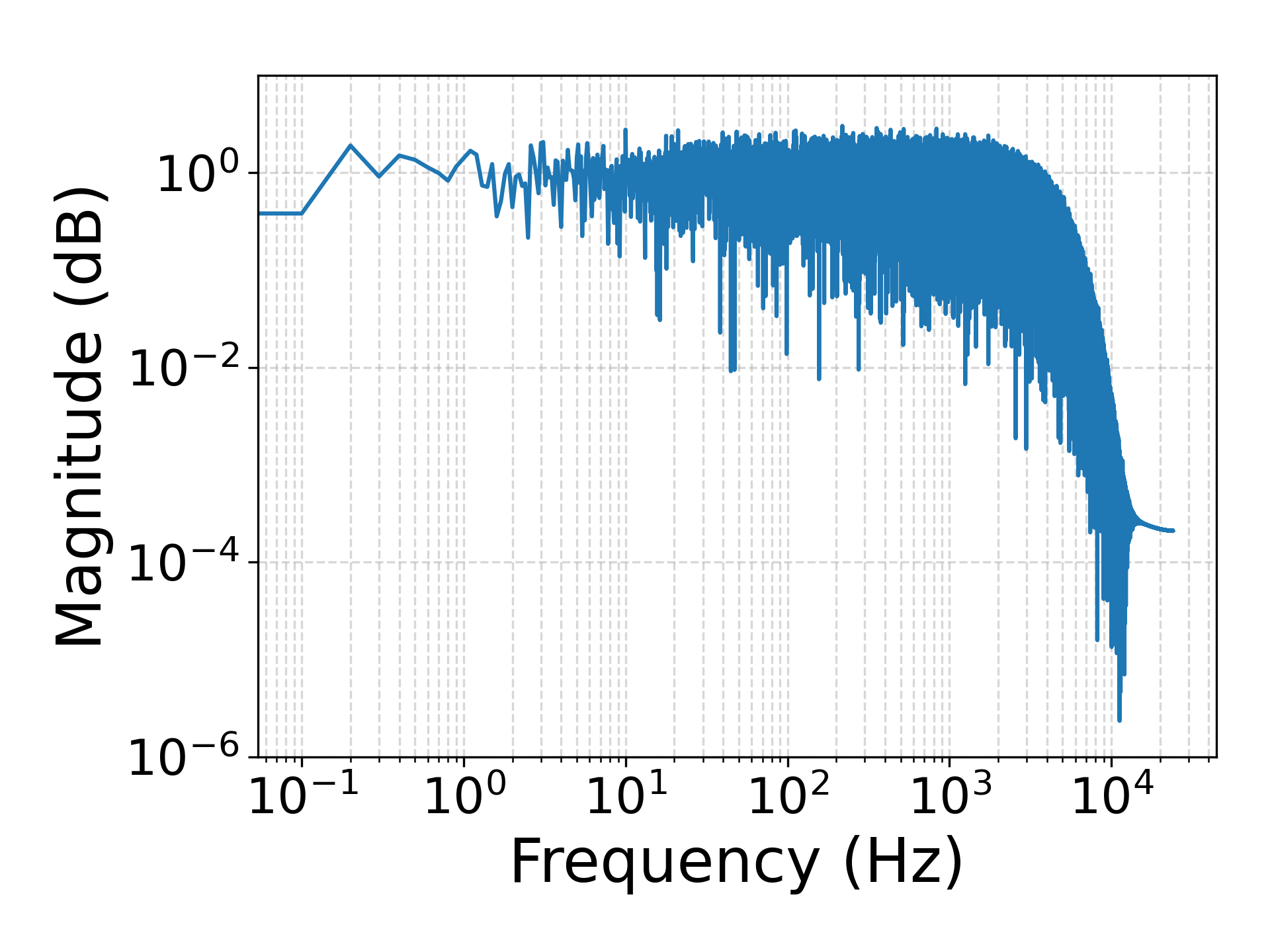}
            \put(-5,65){\colorbox{white}{\textbf{f)}}}
        \end{overpic}
    \end{minipage}

    \caption{White noise signal sound source at $434\mathrm{\,m}$ from the microphone. a) spectrogram and d) FFT of the audio generated with \textit{pyroomacoustics}. b) spectrogram and e) FFT of the audio generated with \textit{pyroadacoustics}. c) spectrogram and f) FFT of the audio generated with \textit{DynamicSound} simulator.}
    \label{fig:static_whitenoise}
\end{figure*}

The second experiment evaluates the behavior of each simulator when modeling a broadband (white-noise) source located at the same distance of $343\mathrm{\,m}$. Similar to the previous case, the received signal begins approximately $1\mathrm{\,s}$ after emission, as expected. The FFT images shown in Fig.~\ref{fig:static_whitenoise}(d, e, f) reveal the characteristic attenuation of high-frequency components due to air absorption. This effect is visible across all simulators, although the degree of attenuation has small changes depending on the specific air-absorption models implemented by each tool.

\subsection{Simulation of a Moving Source}

The third experiment investigates the ability of the simulators to model dynamic scenarios involving source motion. Among the evaluated tools, only \textit{pyroadacoustics} and the proposed \textit{DynamicSound} framework support moving sources. In this scenario, the source initially remains at low constant velocity of $0.5\mathrm{\,m/s}$ for the first three seconds and then starts to move at $49\mathrm{\,m/s}$ toward the microphone. We use these conditions as one of the limitations of \textit{pyroadacoustics} is that it does not allow a source to remain stationary for an initial period before starting to move, whereas the proposed \textit{DynamicSound} framework provides greater flexibility, enabling the simulation of trajectories with intervals of stationary behavior.

\begin{figure*}[!htp]
    \centering

    \begin{minipage}{0.32\textwidth}
        \centering
        \begin{overpic}[width=\linewidth]{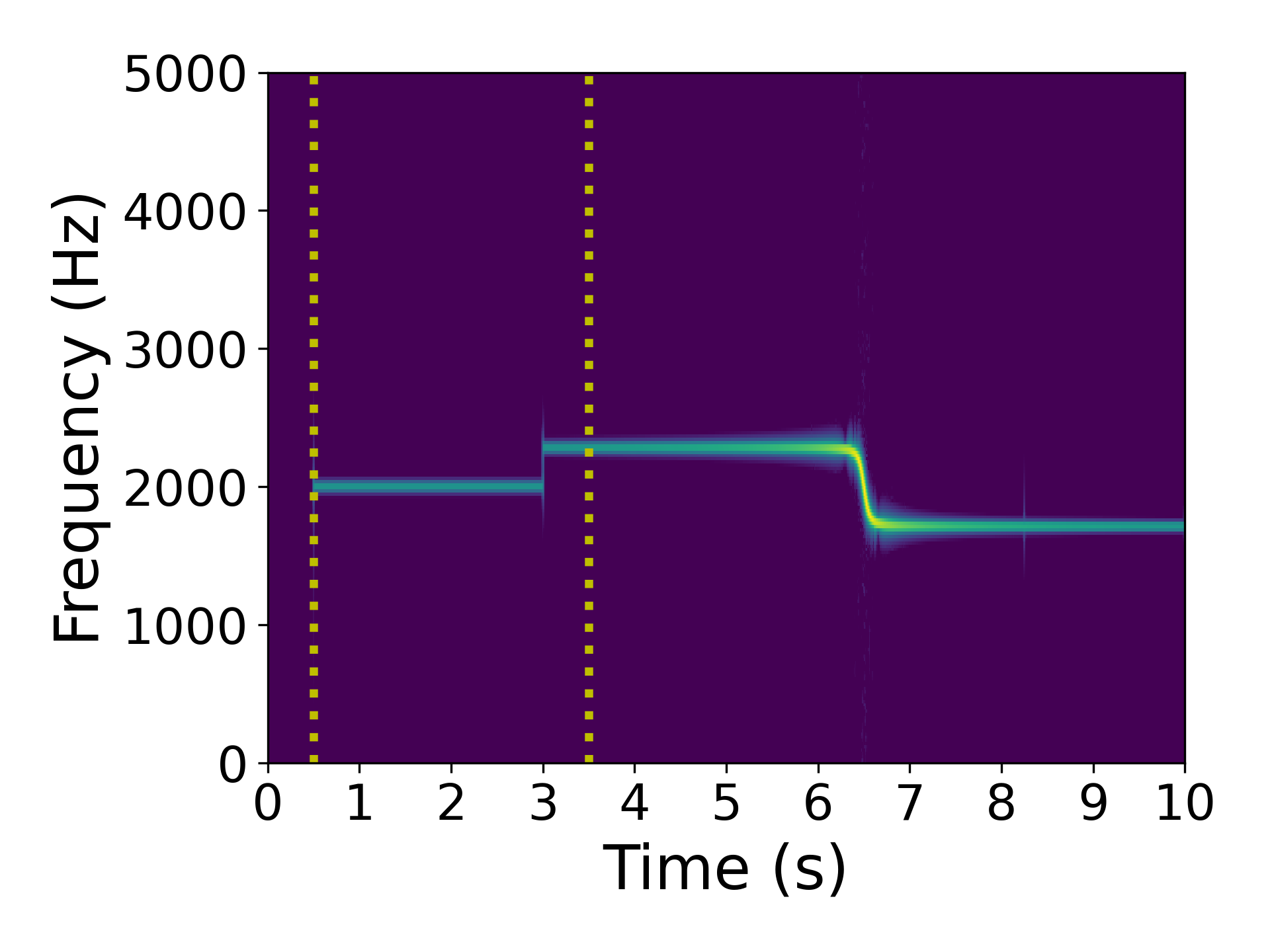}
            \put(-5,65){\colorbox{white}{\textbf{a)}}}
        \end{overpic}
    \end{minipage}\hfill
    \begin{minipage}{0.32\textwidth}
        \centering
        \begin{overpic}[width=\linewidth]{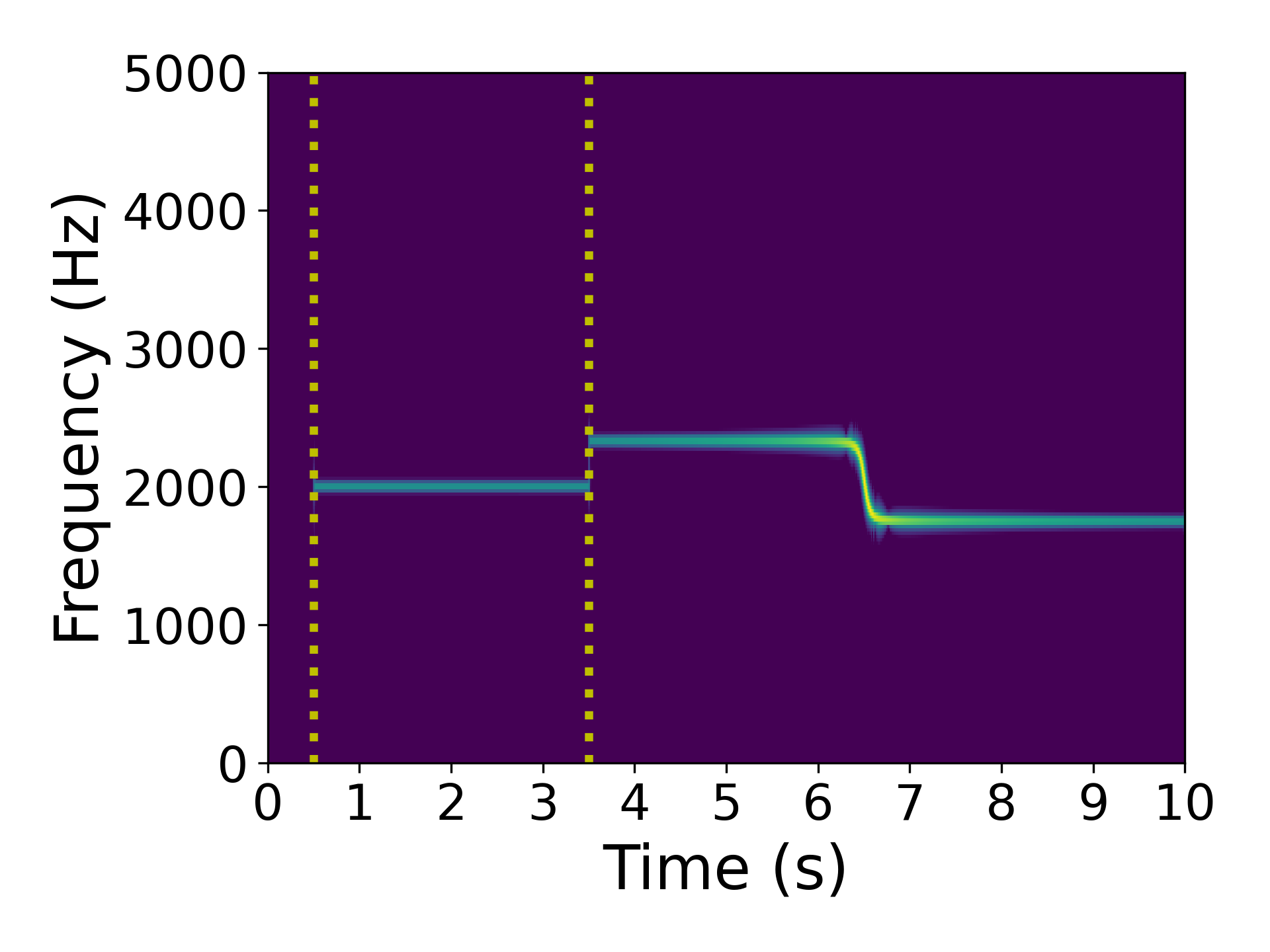}
            \put(-5,65){\colorbox{white}{\textbf{b)}}}
        \end{overpic}
    \end{minipage}\hfill
    \begin{minipage}{0.32\textwidth}
        \centering
        \begin{overpic}[width=\linewidth]{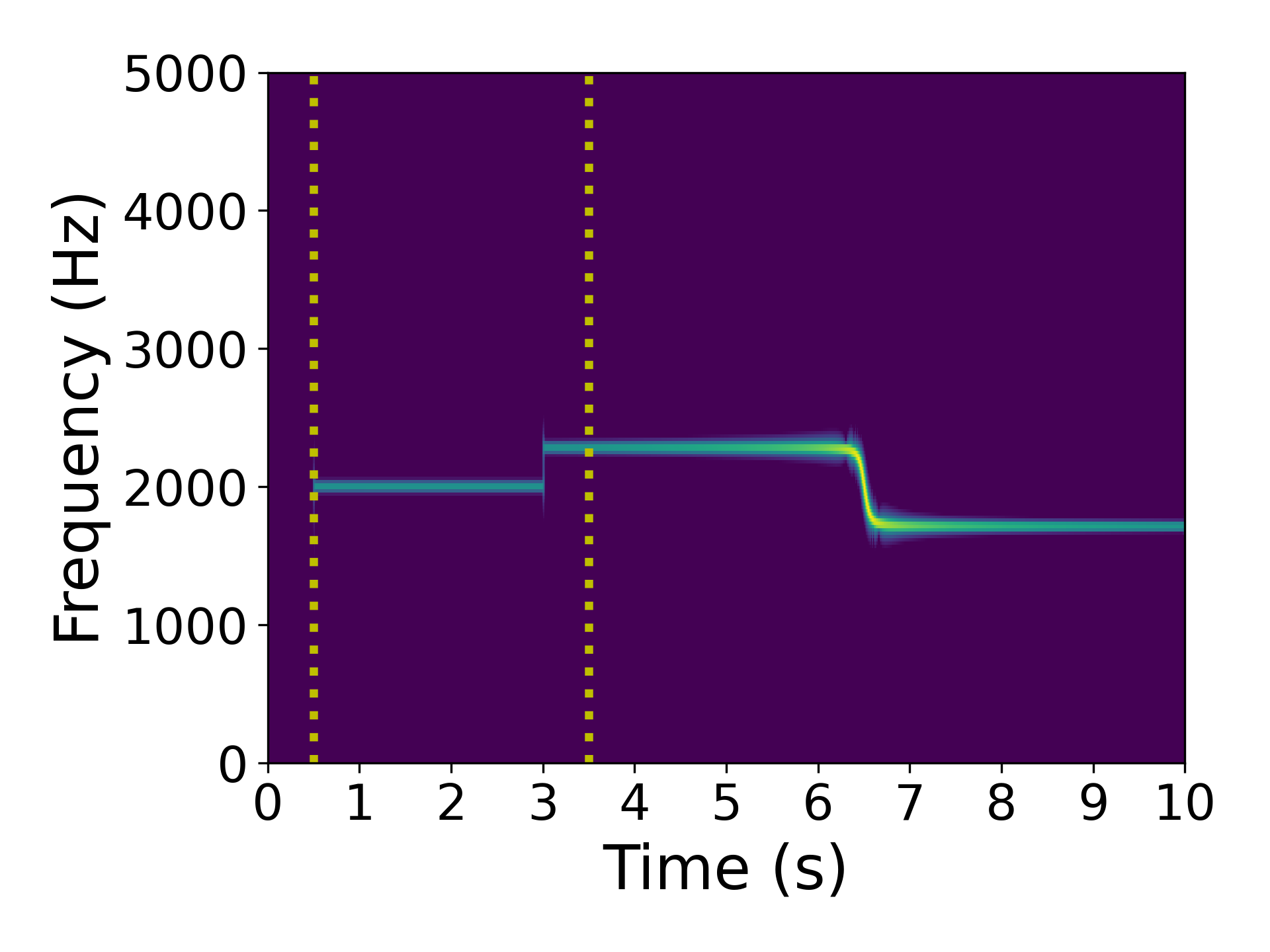}
            \put(-5,65){\colorbox{white}{\textbf{c)}}}
        \end{overpic}
    \end{minipage}

    \vspace{0em}

    \begin{minipage}{0.32\textwidth}
        \centering
        \begin{overpic}[width=\linewidth]{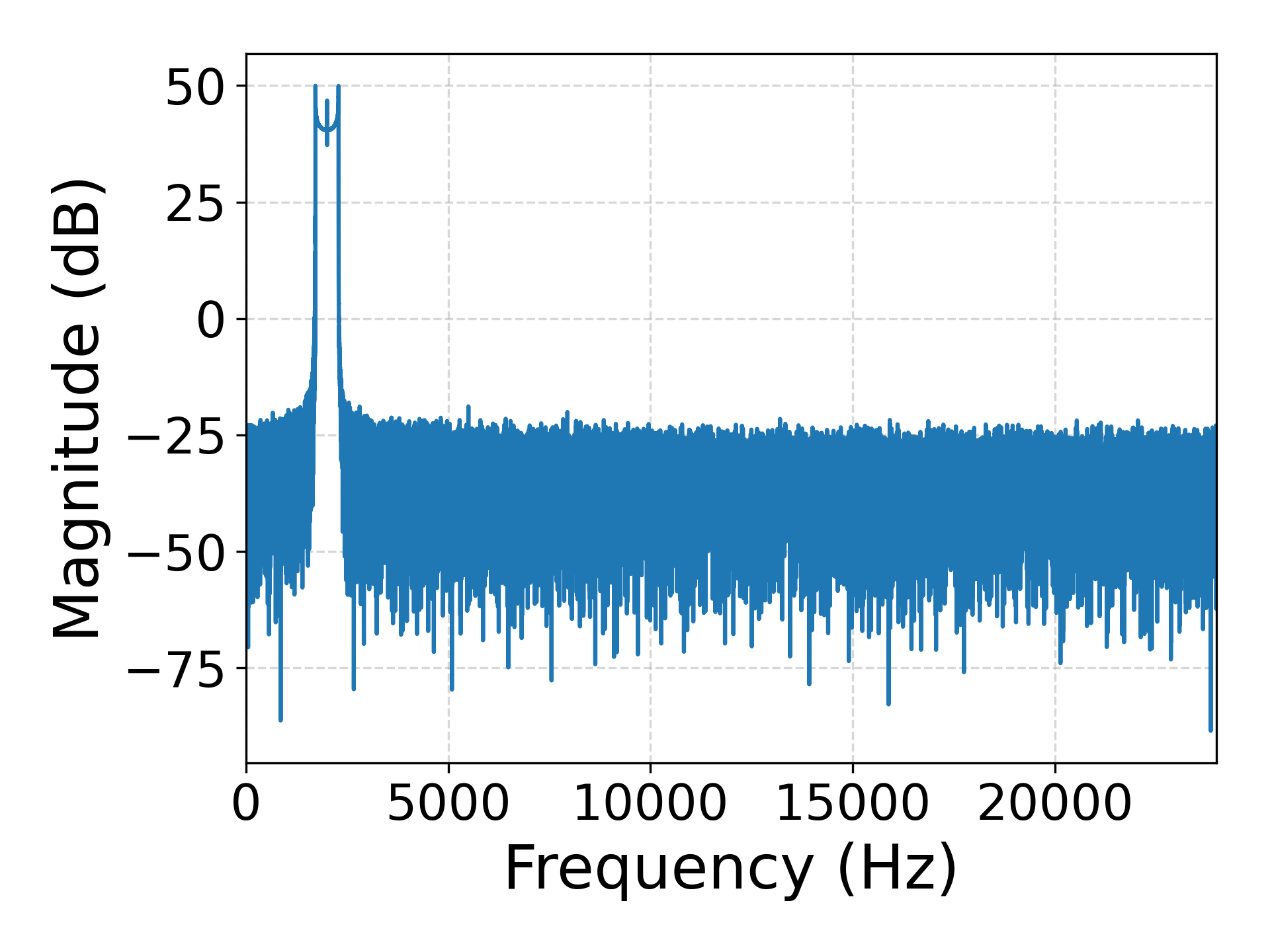}
            \put(-5,65){\colorbox{white}{\textbf{d)}}}
        \end{overpic}
    \end{minipage}\hfill
    \begin{minipage}{0.32\textwidth}
        \centering
        \begin{overpic}[width=\linewidth]{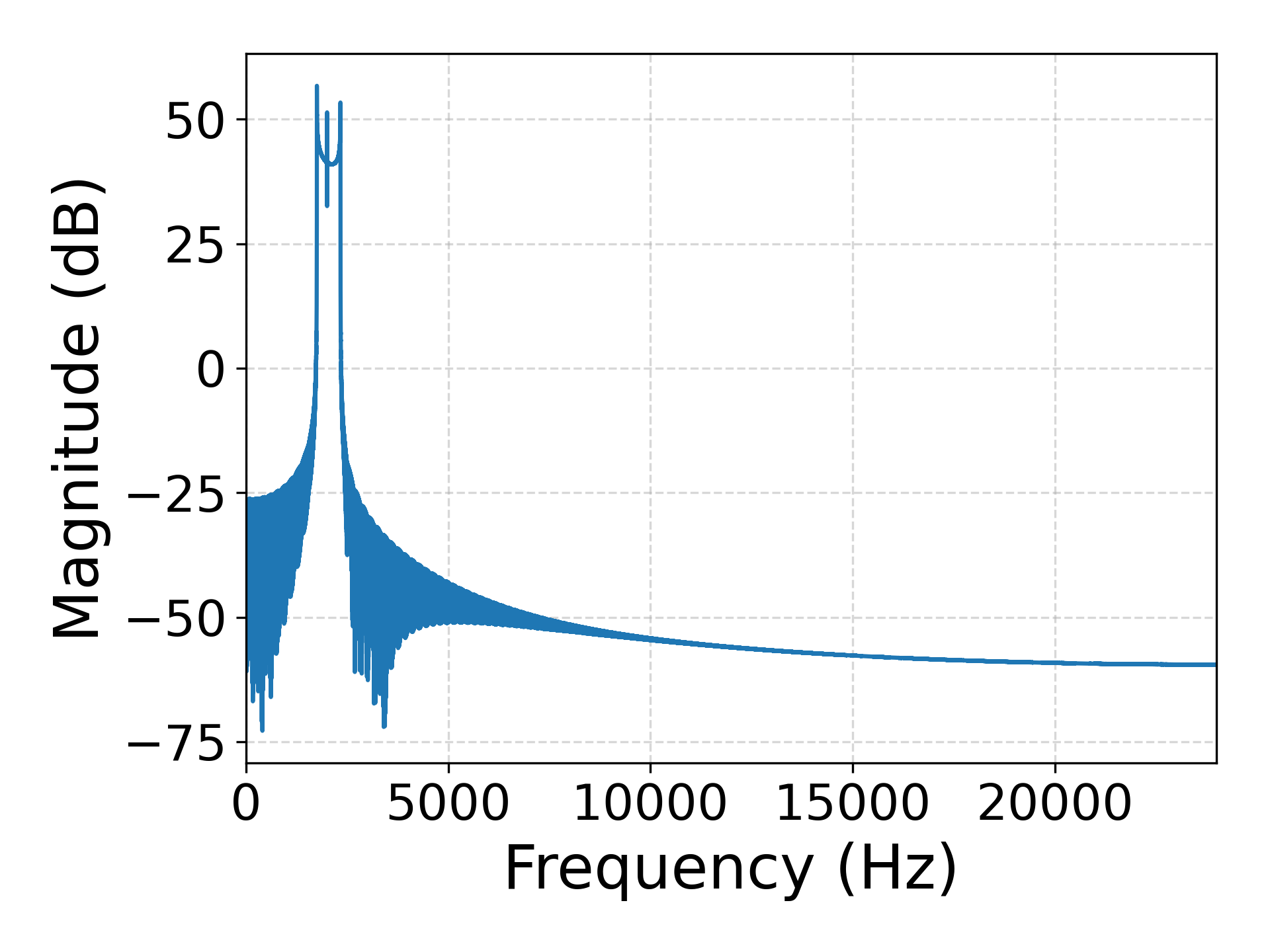}
            \put(-5,65){\colorbox{white}{\textbf{e)}}}
        \end{overpic}
    \end{minipage}\hfill
    \begin{minipage}{0.32\textwidth}
        \centering
        \begin{overpic}[width=\linewidth]{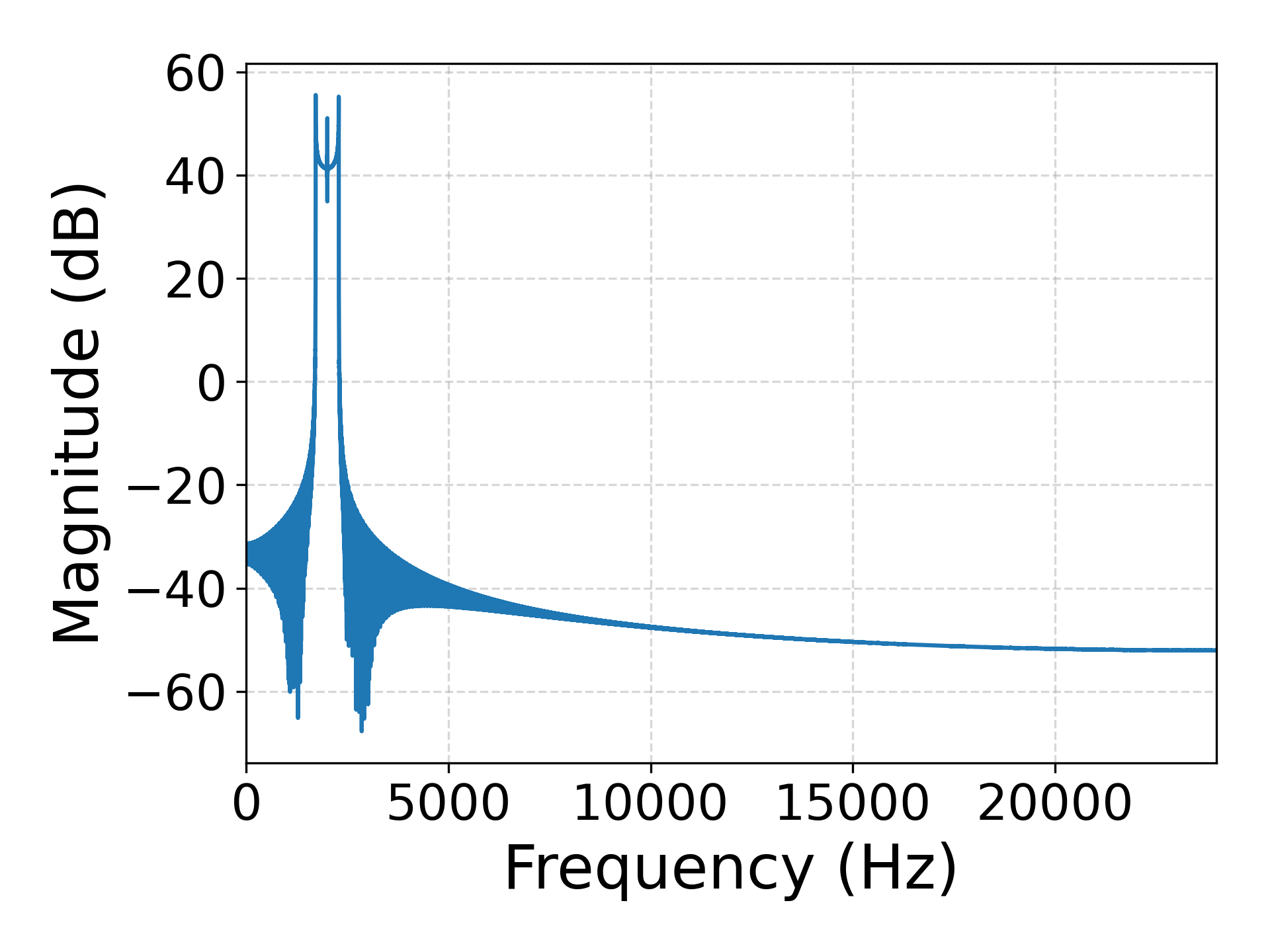}
            \put(-5,65){\colorbox{white}{\textbf{f)}}}
        \end{overpic}
    \end{minipage}

    \caption{Sinusoidal sound source at \(151m\) from the microphone, starts to emit the signal at time $0\mathrm{s}$ moving slowly towards the microphone for 3 seconds with a $speed=0.5\mathrm{m/s}$, then starts moving faster from the point $(171.5, 3 ,1)$ to $(-171.5, 3, 1)$ with a speed of $49\mathrm{\,m/s}$. a) spectrogram and d) FFT of the audio generated with \textit{pyroadacoustics}. b) spectrogram and e) FFT of the audio generated with \textit{DynamicSound} simulator. c) spectrogram and f) FFT of the audio generated with \textit{DynamicSound} simulator inverting the paths of the source and the microphone.}
    \label{fig:dynamic_sin}
\end{figure*}

In this scenario, the source is positioned at an initial distance of $171.5\mathrm{\,m}$, resulting in an expected propagation delay of approximately $0.5\mathrm{\,s}$. As illustrated in Fig.~\ref{fig:dynamic_sin}, both simulators correctly reproduce this delay, with the first noticeable signal arrival occurring at $\mathrm{t} = 0.5$\,s (first vertical marker). However, differences emerge when the source changes speed at $\mathrm{t} = 3$\,s. In \textit{pyroadacoustics}, the change in motion is reflected immediately at the receiver as visible in Fig.~\ref{fig:dynamic_sin}(a), whereas in \textit{DynamicSound} the corresponding effect appears at $\mathrm{t} = 3.5$\,s as shown in Fig.~\ref{fig:dynamic_sin}(b) (second vertical marker), this behavior is physically correct, as it reflects the finite propagation time of the emitted wavefronts.
Figures~\ref{fig:dynamic_sin}(c, f) show the simulation performed with \textit{DynamicSound} by inverting the paths of the source and the microphone. In this configuration, the microphone is moving while the source remains stationary, yielding results that match those obtained with \textit{pyroadacoustics} in Figs.~\ref{fig:dynamic_sin}(a, d). This highlights that \textit{pyroadacoustics} actually models array motion rather than true source motion, leading to physically incorrect behavior if a simulation of a moving source is needed.

Frequency-domain analysis further highlights differences between the frameworks. As shown in Fig.~\ref{fig:dynamic_sin}, both simulators emphasize the three main frequency components of the source corresponding to the three different relative speeds of the sound source, but \textit{DynamicSound} exhibits reduced high-frequency noise contamination, indicating a more stable and artifact-free modeling of the Doppler-induced modulation during motion.


\subsection{Moving White Noise Source with Ground Reflection}

In this experiment, we analyze the behavior of the simulators in a dynamic scenario where a broadband source moves horizontally above a reflective plane. The source emits a white-noise signal while traveling in a linear trajectory at a constant speed of $5\mathrm{\,m/s}$ along the trajectory from $p_A=(-20,\,3,\,1)$\,m to $p_B=(20,\,3,\,1)$\,m, where positions are given in Cartesian coordinates $(x, y, z)$. The microphone is positioned at $(0,\,0,\,1)$\,m, 1\,m above the terrain, resulting in a configuration where both the direct and the ground-reflected waves reach the receiver.

\begin{figure*}[!htp]
    \centering

    \begin{minipage}{0.32\textwidth}
        \centering
        \begin{overpic}[width=\linewidth]{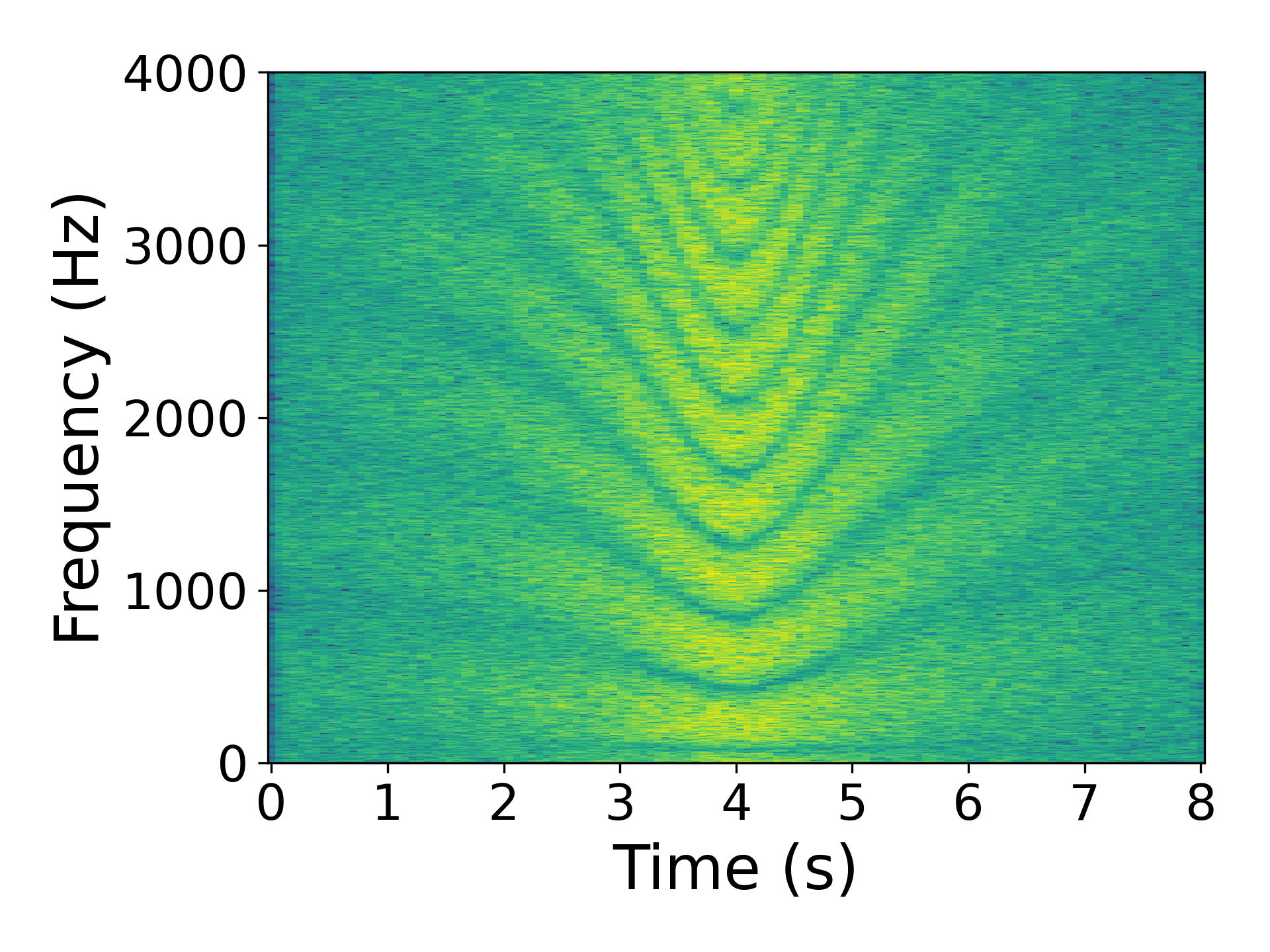}
            \put(-5,65){\colorbox{white}{\textbf{a)}}}
        \end{overpic}
    \end{minipage}\hfill
    \begin{minipage}{0.32\textwidth}
        \centering
        \begin{overpic}[width=\linewidth]{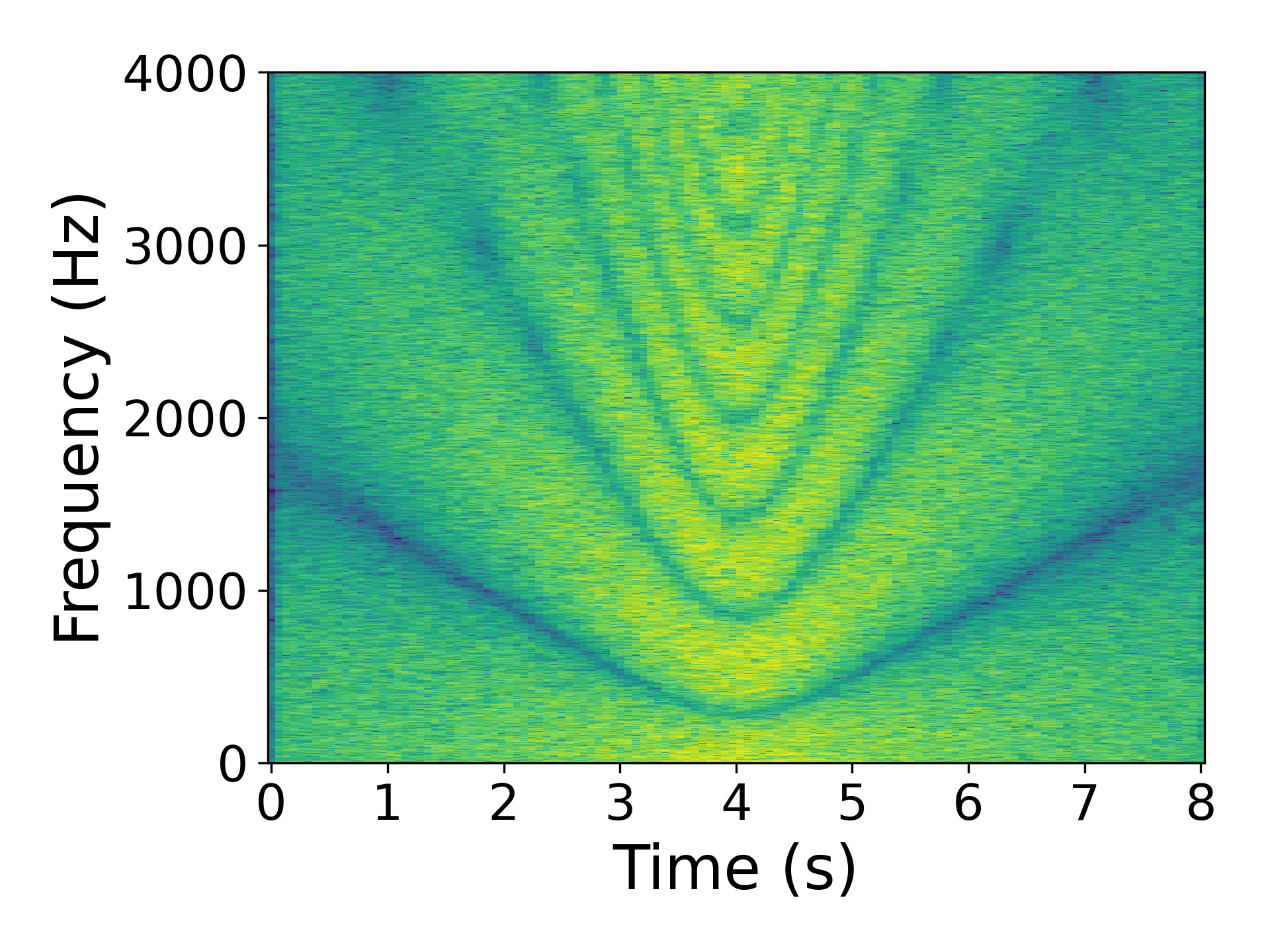}
            \put(-5,65){\colorbox{white}{\textbf{b)}}}
        \end{overpic}
    \end{minipage}\hfill
    \begin{minipage}{0.32\textwidth}
        \centering
        \begin{overpic}[width=\linewidth]{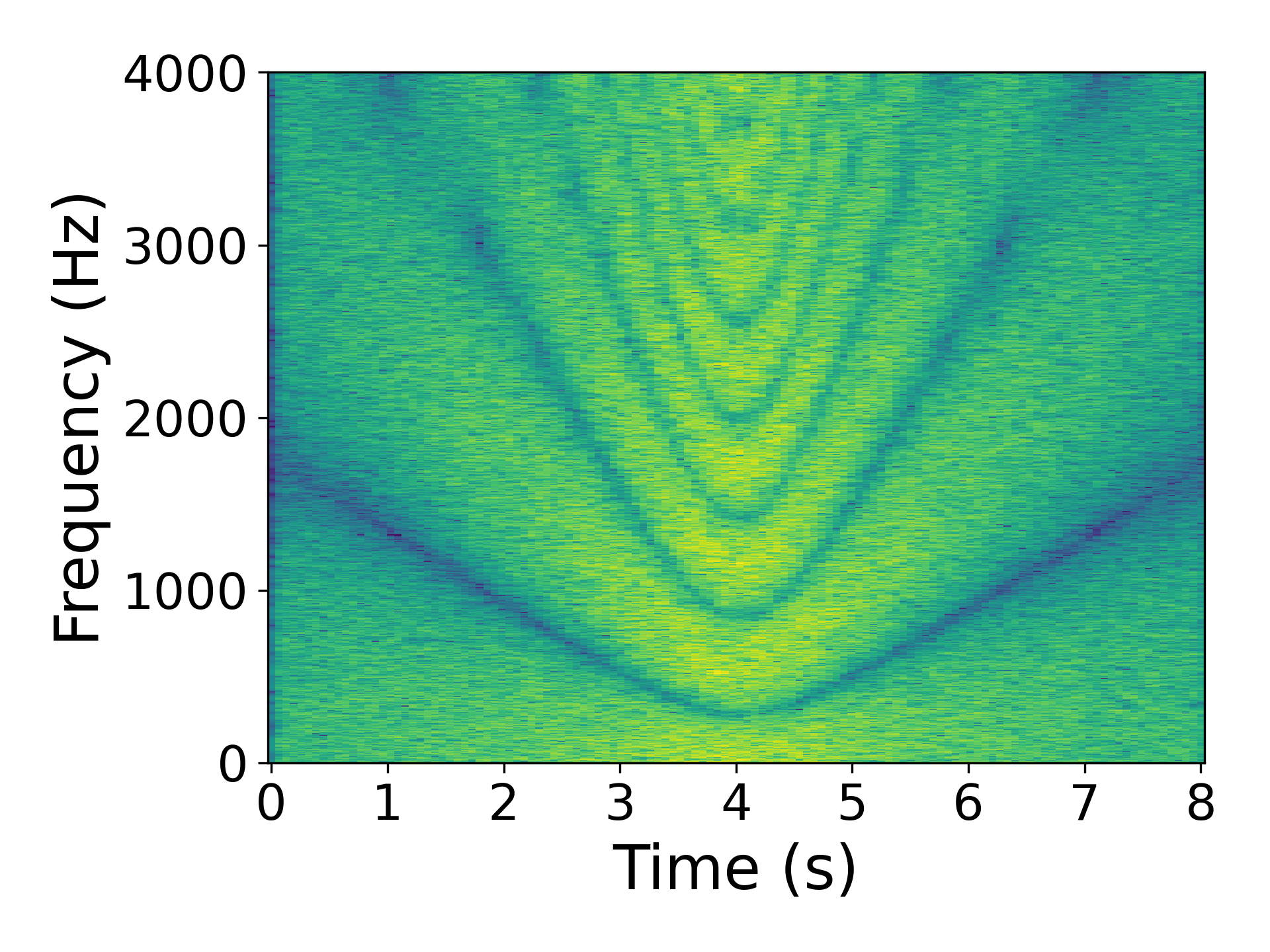}
            \put(-5,65){\colorbox{white}{\textbf{c)}}}
        \end{overpic}
    \end{minipage}

    \vspace{0em}

    \begin{minipage}{0.32\textwidth}
        \centering
        \begin{overpic}[width=\linewidth]{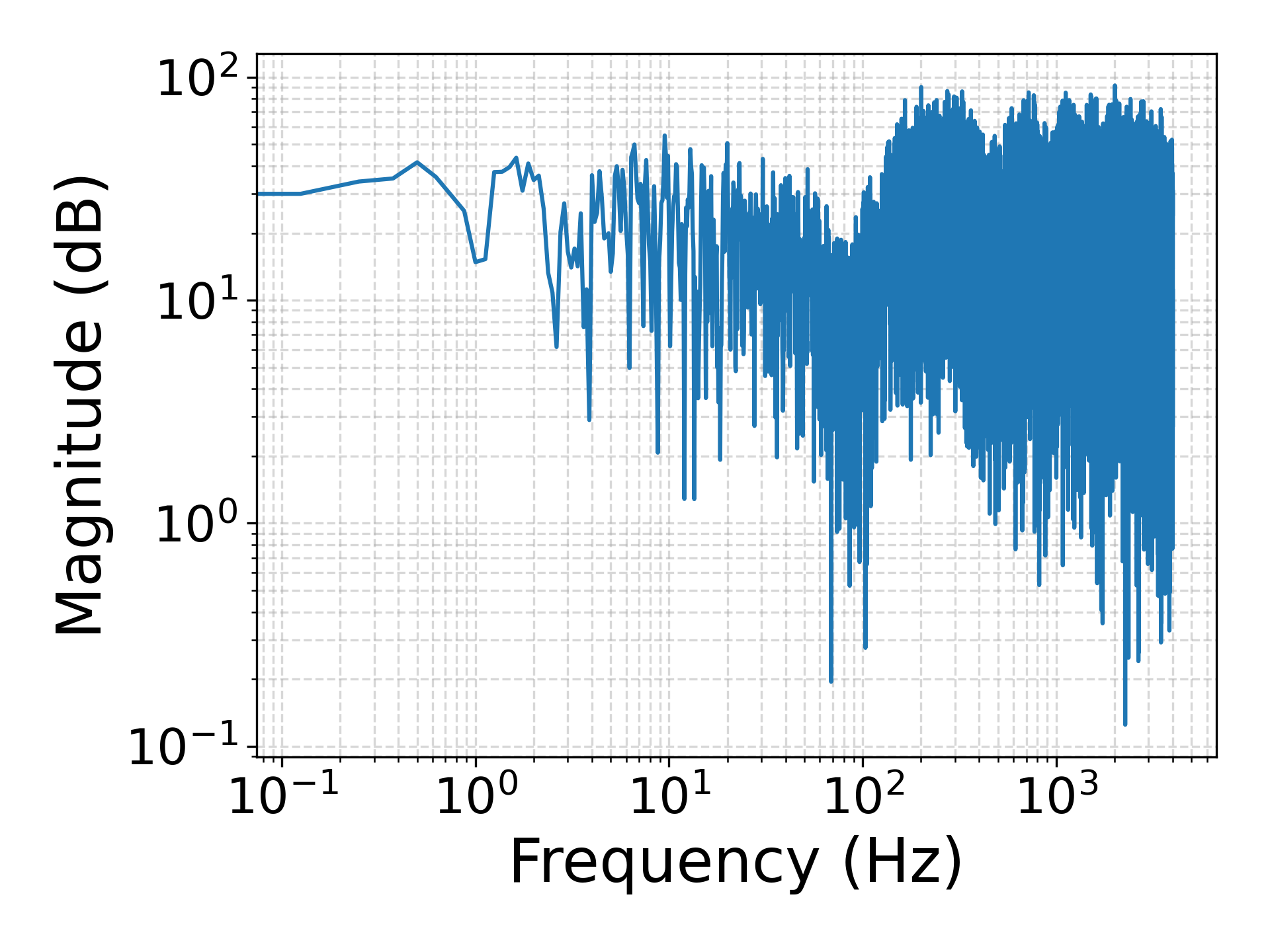}
            \put(-5,65){\colorbox{white}{\textbf{d)}}}
        \end{overpic}
    \end{minipage}\hfill
    \begin{minipage}{0.32\textwidth}
        \centering
        \begin{overpic}[width=\linewidth]{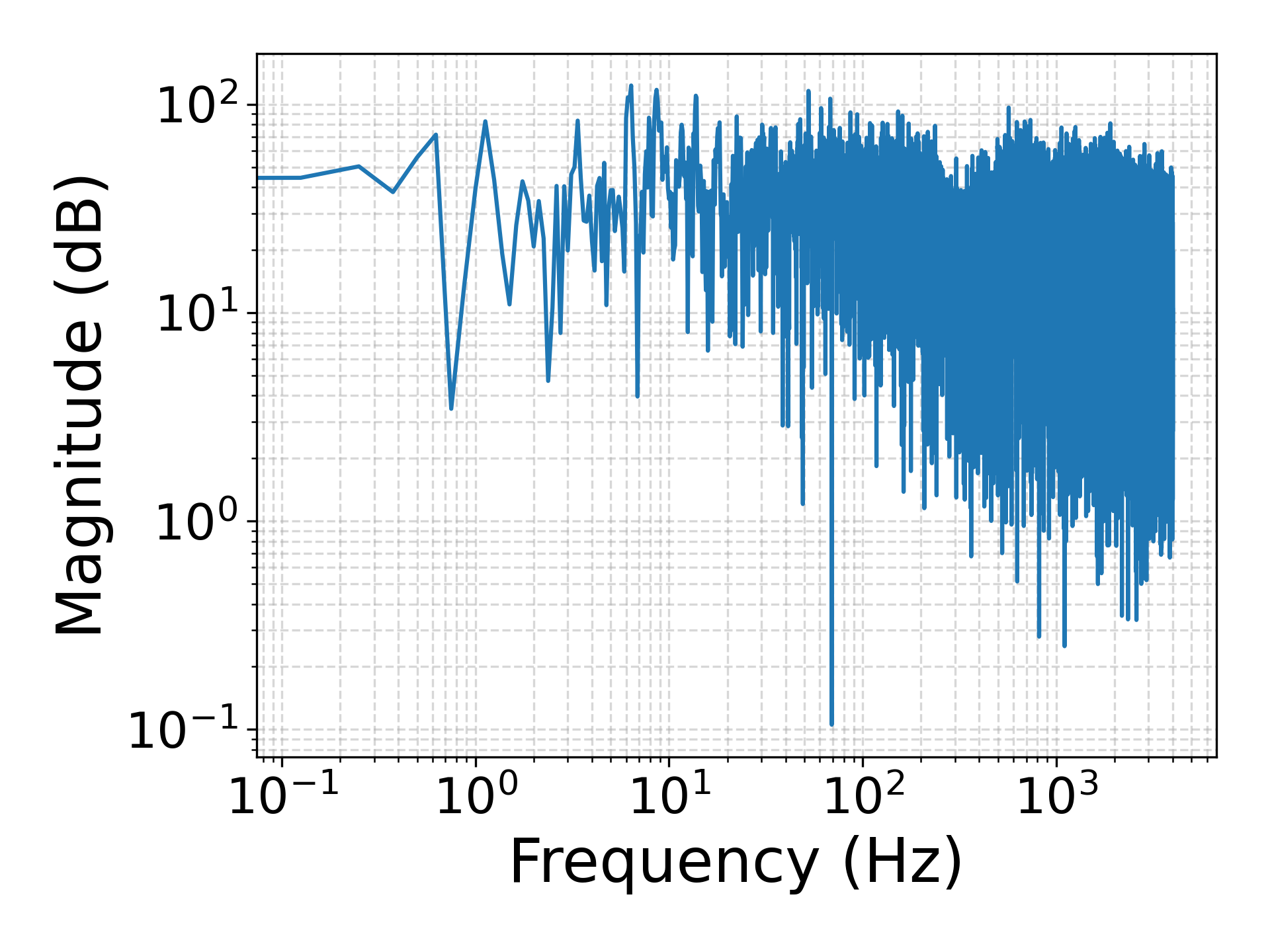}
            \put(-5,65){\colorbox{white}{\textbf{e)}}}
        \end{overpic}
    \end{minipage}\hfill
    \begin{minipage}{0.32\textwidth}
        \centering
        \begin{overpic}[width=\linewidth]{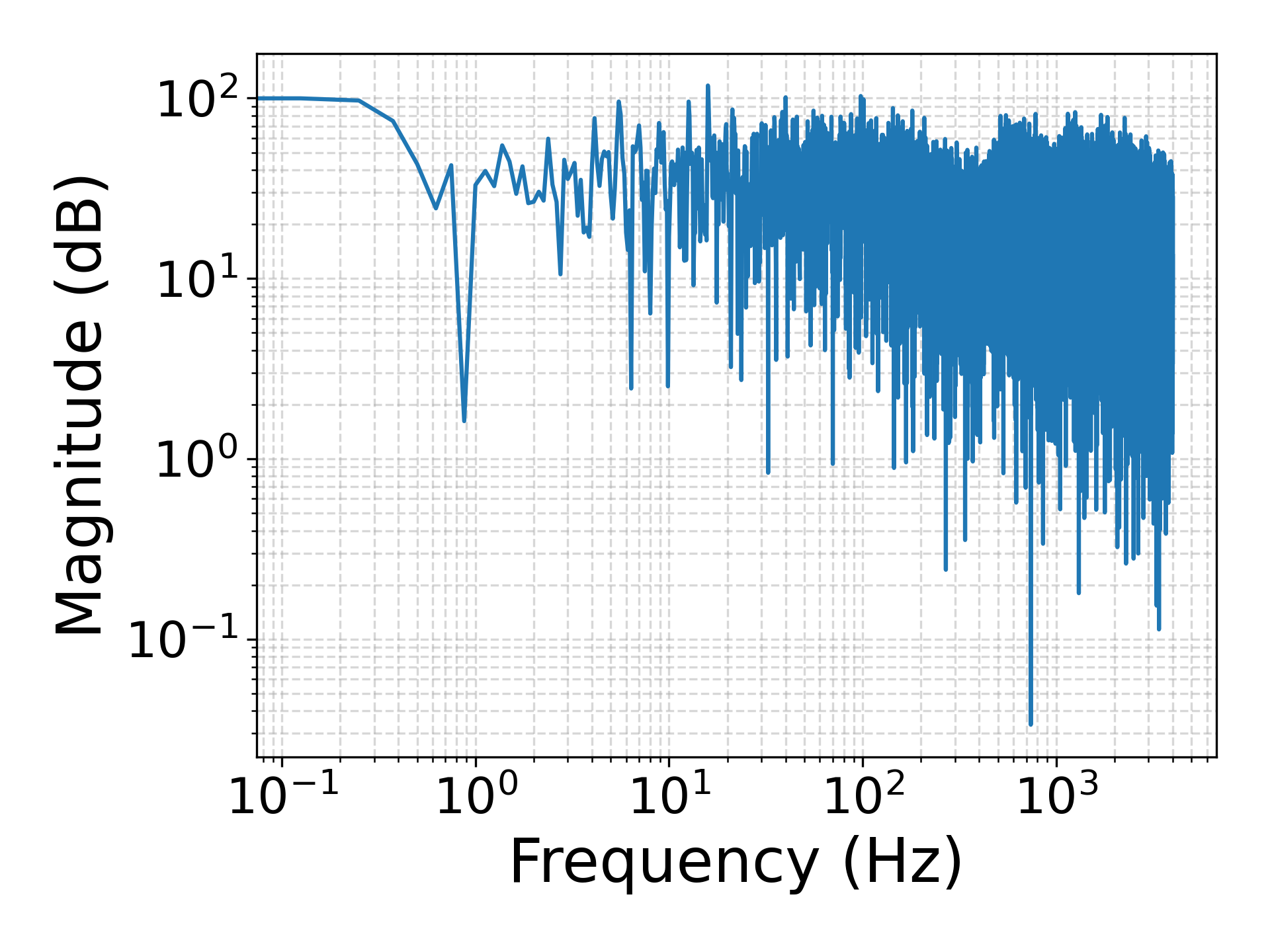}
            \put(-5,65){\colorbox{white}{\textbf{f)}}}
        \end{overpic}
    \end{minipage}

    \caption{White noise starts to emit the signal at time $0\mathrm{s}$ moving from the point $(-20,\,3,\,1)\mathrm{\,m}$ to $(20,\,3,\,1)\mathrm{\,m}$ with a speed of $5\mathrm{\,m/s}$. a) spectrogram and d) FFT of the audio generated with \textit{pyroadacoustics}. b) spectrogram and c) FFT of the audio generated with \textit{DynamicSound} simulator. c) spectrogram and f) FFT of the audio generated with \textit{DynamicSound} simulator inverting the paths of the source and the microphone.}
    \label{fig:dynamic_white}
\end{figure*}

As shown in Fig.~\ref{fig:dynamic_white}, the combination of the direct path and its reflection generates a characteristic comb-filter pattern that varies over time as the source moves. This effect arises from constructive and destructive interference between the two propagation paths, whose relative phases change continuously with the source–receiver geometry. The spectrogram clearly highlights this dynamic modulation, with periodic spectral nulls shifting as the path-length difference evolves.

A noticeable discrepancy emerges between the spectrograms produced by the two simulators. In this case, inverting the trajectories of the source and the microphone in the \textit{DynamicSound} simulation yields the results shown in Figs.~\ref{fig:dynamic_white}(c, f), which remains largely consistent with those of the previous scenario in Figs.~\ref{fig:dynamic_white}(b, e). This indicates that the observed differences are primarily attributable to the distinct reflection models implemented in each tool, rather than to differences in the handling of propagation delays.
In \textit{pyroadacoustics}, the reflection coefficient depends on both the acoustic absorption of the material and the angle of incidence of the incoming wave, resulting in a frequency and angle dependent attenuation of the reflected component. In contrast, the current implementation of \textit{DynamicSound} uses a fully reflective surface model as a simplifying assumption, with no angle-dependent energy loss.

\subsection{Vehicle-Tracking Scenario Using a Microphone Array}

\begin{figure}[!ht]
    \centering
    \includegraphics[width=0.85\linewidth]{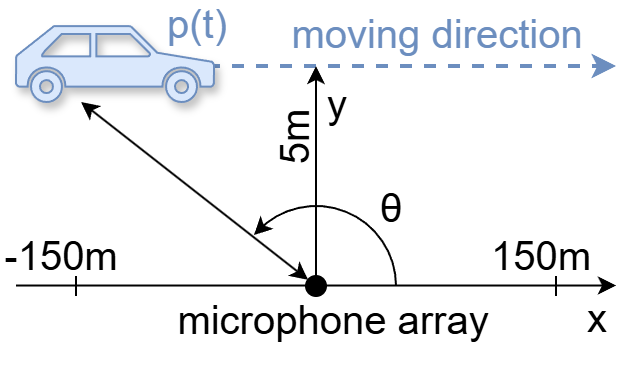}
    \caption{Simulated scenario of a moving car along the road relative to the microphone array, showing the vehicle trajectory and the array position used for DOA estimation.}
    \label{fig:car_scenario}
\end{figure}

The final experiment replicates the scenario studied in \cite{zhang2014design}, where a compact microphone array is used to estimate the direction of arrival (DOA) of a passing vehicle. In this simulation, an audio recording of a car running off-road is used as the source signal. The vehicle trajectory is modeled as a straight line extending from $-150\mathrm{\,m}$ to $150\mathrm{\,m}$ along the $x$-axis, while the microphone array is positioned $5\mathrm{\,m}$ away from this path as shown in Fig.~\ref{fig:car_scenario}. The source moves at a constant speed of 15\,m/s equivalent to 54\,km/h forming an angle $\theta$ with the microphone array described by the equation \ref{eq:angle}, where $p_x(t)$ is the car position along the moving direction and $p_y=5\mathrm{\,m}$ is the distance of the microphone from the trajectory.

\begin{equation}
    \theta = \frac{180}{\pi} \cdot \left( \frac{\pi}{2} + \mathrm{arctan} \left( \frac{-p_x(t)}{p_y} \right) \right)
    \label{eq:angle}
\end{equation}

\begin{figure}[!ht]
    \centering
    \includegraphics[width=\linewidth]{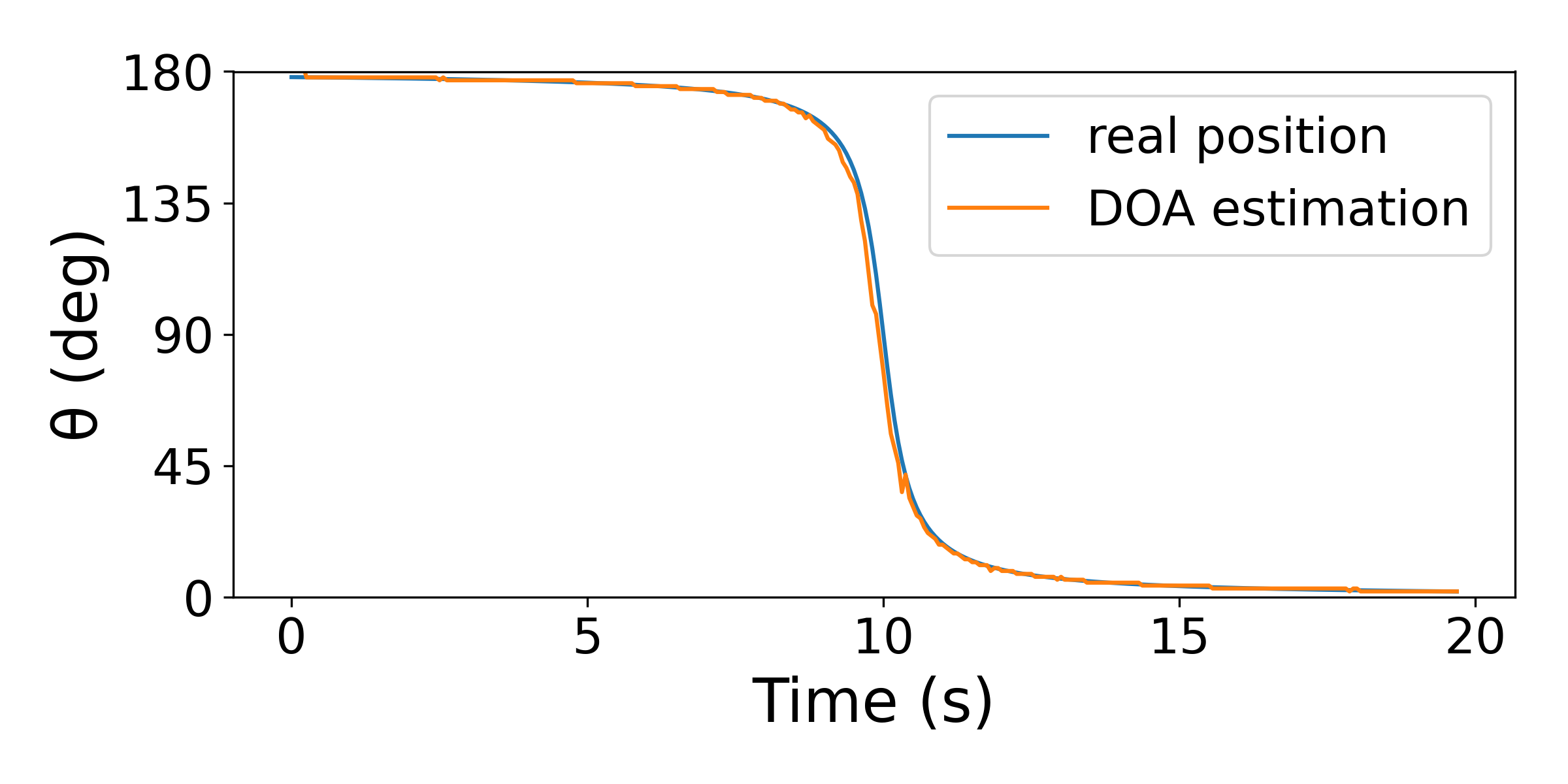}
    \caption{Diirection of arrival $\theta$ estimated using the NormMUSIC algorithm compared with the real angle position.}
    \label{fig:doa_car}
\end{figure}

The array geometry follows the configuration as shown in \cite{zhang2014design}, consisting of four microphones arranged in a circular pattern with a diameter of 4\,cm. This compact layout enables high spatial resolution despite the small aperture, making it appropriate for DOA estimation of moving sources.

As shown in Fig.~\ref{fig:doa_car}, the signals generated by \textit{DynamicSound} can be directly processed using the NormMUSIC \cite{salvati2014incoherent} algorithm implementation available in the \textit{pyroomacoustics} library \cite{scheibler2018pyroomacoustics}. The resulting DOA estimates track the motion of the simulated vehicle as it approaches, passes, and moves away from the array. This demonstrates that the simulator not only reproduces realistic multichannel recordings but also integrates smoothly with standard spatial-audio processing pipelines, enabling evaluation of array-processing algorithms in controlled and repeatable conditions.


\section{Conclusions}
\label{sec:conclusions}

This work introduced \textit{DynamicSound}, an open-source acoustic simulation framework designed to generate realistic multichannel audio for static and moving sound sources in open-field environments. The simulator incorporates key physical effects—time-of-flight delays, Doppler shifts, geometric spreading, and air absorption—while supporting flexible microphone-array configurations and source trajectories. By being openly released to the community, DynamicSound promotes transparency, reproducibility, and extensibility, enabling researchers and practitioners to inspect, modify, and build upon the proposed methods. The resulting system provides a valuable tool for developing and testing spatial-audio algorithms, beamforming strategies, and machine-learning models for sound classification and localization.

The experiments conducted in this study demonstrate several important outcomes. First, in static long-distance scenarios, \textit{DynamicSound} reproduces physically correct propagation delays and spectral behavior, matching the expected responses derived from acoustic theory and aligning consistently with existing open-source simulators. Second, in broadband and reflective conditions, the simulator accurately models frequency-dependent attenuation and interference patterns such as time-varying comb filtering. Although the current reflection model assumes perfectly reflective surfaces, the overall qualitative behavior remains consistent with the underlying physics and highlights opportunities for future improvement.

Most notably, \textit{DynamicSound} proves capable of handling dynamic scenes involving moving sources, reproducing time delays and Doppler-related effects with correct temporal causality. 
Unlike the \textit{pyroadacoustics} tool that implements instantaneous motion updates, the proposed simulator preserves the finite propagation time between emission and reception, ensuring realistic synchronization between source movement and observed signal characteristics. This advantage becomes evident in scenarios with abrupt velocity changes, where \textit{DynamicSound} produces a physically accurate delayed response.

Finally, the vehicle-tracking experiment using a compact microphone array demonstrates that the simulated data integrates seamlessly with established array-processing algorithms such as NormMUSIC, enabling reliable DOA estimation throughout the motion of the source. This confirms the simulator’s suitability for research on real-world applications including traffic monitoring, robotic perception, drone detection, and autonomous navigation.

Overall, \textit{DynamicSound} provides a flexible, extensible, and physically grounded framework for multichannel sound simulation. Future extensions will focus on incorporating more advanced reflection models, diffraction and occlusion effects, and the ability to simulate complex environments with multiple interacting surfaces. These enhancements will further expand the applicability of the simulator to indoor robotics, acoustic scene reconstruction, and high-fidelity virtual acoustics.

\section{funding}

This publication is part of the project PNRR-NGEU which has received funding from the MUR – DM 352/2022

\bibliographystyle{IEEEtran}

\begin{IEEEbiography}[{\includegraphics[width=1in,height=1.25in,clip,keepaspectratio]{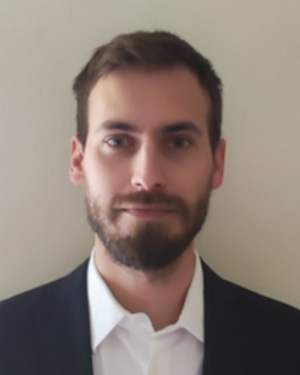}}]{Luca Barbisan}
received the bachelor’s degree in electronics and telecommunication engineering from University of Trento, Italy, in 2017 and the master's degree in electronics engineering from Politecnico di Torino, Italy, in 2021. He is currently working toward the Ph.D. degree in electronics and telecommunications engineering at Politecnico di Torino. His current research interests include digital signal processing and modern machine learning techniques, mainly applied to audio signals.
\end{IEEEbiography}

\begin{IEEEbiography}[{\includegraphics[width=1in,height=1.25in,clip,keepaspectratio]{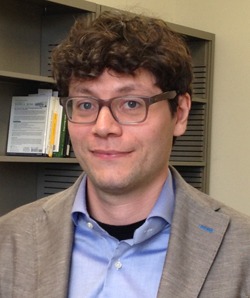}}]{Marco Levorato}
received the PhD degree in electrical engineering from the University of Padova, Italy, in 2009. He is a professor with the Computer Science Department, University of California, Irvine. Between 2010 and 2012, he was a postdoctoral researcher jointly at Stanford and the University of Southern California. His research interests include focused on distributed computing over unreliable wireless systems, especially for autonomous vehicles and robotic applications.

\end{IEEEbiography}

\begin{IEEEbiography}[{\includegraphics[width=1in,height=1.25in,clip,keepaspectratio]{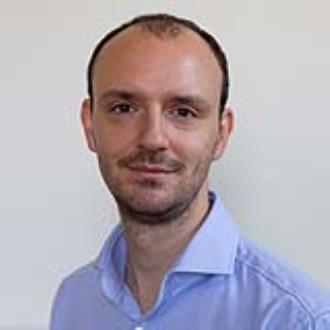}}]{Fabrizio Riente}
received the M.Sc. degree (magna cum laude) in electronic engineering and the Ph.D. degree from Politecnico di Torino, Turin, Italy, in 2012 and 2016, respectively. In 2016 and 2019, he was a Postdoctoral Research Associate with the Technical University of Munich, Munich, Germany. He is currently an Assistant Professor with the Department of Electronics and Telecommunications, Politecnico di Torino. His research interests include device and compact modeling and circuit design for beyond-CMOS and nano-computing systems, with emphasis on field-coupled nanotechnologies. His current research also spans embedded and low-power signal processing, including multichannel audio and acoustic sensing for microphone arrays, as well as energy-efficient IoT and edge-computing platforms.
\end{IEEEbiography}


\vfill

\end{document}